\documentclass[compress,sort,twocolumn]{autart-small-skip}
\usepackage{cite}

\journal{Elsevier}
\usepackage{graphicx}          
\usepackage{subfigure}
\usepackage{fancyhdr}
\usepackage{amsmath}
\usepackage{multirow}
\usepackage{amssymb }
\usepackage{color}
\usepackage{graphics} 
\usepackage{graphicx}

\usepackage{amsmath}
\newtheorem{theorem}{\textbf{Theorem}}
\newtheorem{lemma}{\textbf{Lemma}}
\newtheorem{example}{\textbf{Example}}
\newtheorem{corollary}{\textbf{Corollary}}
\newtheorem{remark}{\textbf{Remark}}
\newtheorem{definition}{\textbf{Definition}}

\newtheorem{proposition}{\textbf{Proposition}}
\newtheorem{assumption}{\textbf{Assumption}}
\usepackage{url}
\usepackage{graphicx}
	\newenvironment{proof}{{{\bf Proof:}}}{\hfill $\square$\par}
	
	\usepackage{multirow} 
	\usepackage{xcolor}
	\usepackage{multirow}
	\usepackage{setspace}
	\usepackage{url}
	\usepackage{subfigure}
	\usepackage{fancyhdr}
	\usepackage{amsmath}
	\usepackage{multirow}
	\usepackage{amssymb }
	\usepackage{color}
	\usepackage{graphics} 
	\usepackage{graphicx}
	\usepackage{tikz}
	\usepackage{subeqnarray}
	\usepackage{cases}
	\usepackage{threeparttable}
	\usepackage{algorithm} 
	\usepackage{algpseudocode}
	 \normalsize
	
\begin{document}
		
		\begin{frontmatter}
			
			\title{\large{Rank Resilience of Pattern Matrices against Structured Perturbations with Applications}} 
			
			\thanks[footnoteinfo]{This paper was not presented at any IFAC
				meeting. This work was supported in part by the
				National Natural Science Foundation of China under Grant 62373059.}

			
		
				\author[author1]{Yuan Zhang},
			\author[author1]{Yuanqing Xia}, and
		\author[author1]{Gang Wang}
		
		
		\address[author1]{School of Automation, Beijing Institute of Technology, Beijing, China}
		
		
		\address{Email: $\emph{zhangyuan14@bit.edu.cn, xia\_yuanqing@bit.edu.cn, gangwang@bit.edu.cn}$} 
			
			
		\begin{abstract} In structured system theory, a pattern matrix is a matrix with entries either fixed to zero or free to take arbitrary numbers. The (generic) rank of a pattern matrix is the rank of {\rm{almost all}} its realizations. The resilience of various system properties  is closely tied to the rank resilience of the corresponding pattern matrices. Yet, existing literature predominantly explores the latter aspect by focusing on perturbations that change the zero-nonzero structure of pattern
matrices, corresponding to link additions/deletions. In this paper, we consider the rank resilience of pattern matrices against structured perturbations that can arbitrarily alter the values of a prescribed set of entries, corresponding to link weight variations. We say a pattern matrix is structurally rank $r$ resilient against a perturbation pattern {if} {\rm{almost all}} realizations of this pattern matrix have a rank not less than $r$ under {\rm{arbitrary complex-valued}} realizations of the perturbation pattern. We establish a generic property in this concept. We then present combinatorial necessary and sufficient conditions for a rectangular pattern matrix to lose full rank under given perturbation patterns. We also generalize them to obtain a sufficient condition and a necessary one for losing an arbitrarily prescribed rank. We finally show our results can be applied to the resilience analysis of various properties of structured (descriptor) systems, including controllability and input-state observability, as well as characterizing zero structurally fixed modes. 

	\end{abstract}

	
\end{frontmatter}

{\small{
\section{Introduction}
Recently, security has become crucial in the control and estimation of cyber-physical systems \cite{fawzi2014secure}. Research has explored the secure functionality of various control and estimation algorithms under internal faults (like disconnections of links or nodes \cite{zhang2020generic}) and external attacks (such as sensor or actuator attacks \cite{fawzi2014secure}). A key challenge is understanding the robustness or resilience of system properties, including stability \cite{FabioFragility2018}, stabilization \cite{zhang2022real}, {controllability, and observability} \cite{commault2008observability,Rahimian2013Structural,zhang2019minimal}, against parameter perturbations caused by faults or attacks.


In network systems, {two main types of parameter perturbations affect interconnections.}  The first type arises from adding or removing links, actuators, or sensors, which we refer to as {{\emph{binary perturbations}}}. These perturbations can change the zero/nonzero patterns of system matrices.  To mitigate their effects, designers often include redundant components like links, actuators, and sensors \cite{Sergio_Pequito_2017_robust}.
{Binary perturbations} closely align with the combinatorial nature of network structures, grounded in structured system theory \cite{generic}.  In this framework, system matrices are modeled by pattern matrices ({a.k.a.} structured matrices),
where entries are either fixed to zero or free to take arbitrary values. This model facilitates combinatorial algorithms for network analysis and optimization {\cite{ramos2024minimum}}, {thereby circumventing rounding errors \cite{Ramos2022AnOO}}. Studies have addressed observability resilience under sensor and/or link removals \cite{commault2008observability,Rahimian2013Structural,zhang2019minimal,zhang2023observability}. {As elucidated in \cite{belabbas2022structural}, the (generic) rank of pattern matrices, defined as the rank of almost all\footnote{Throughout this paper, ``almost all'' refers to all parameter values except those within some proper algebraic variety in the corresponding parameter space. Here, a proper algebraic variety is the solution set of a system of nontrivial polynomial equations and has Lebesgue measure zero \cite{Hartshorne2013}.} realizations of the pattern matrices \cite{Murota_Book}, is crucial in various properties of structured systems, including
controllability, input-state observability,  pole assignability in decentralized control, network passivation, and so on {\cite{generic,commault2023dilation}}.} Given this, the resilience analysis conducted in the aforementioned work \cite{commault2008observability,Rahimian2013Structural,zhang2019minimal,belabbas2022structural,zhang2023observability} is closely linked to the rank resilience of pattern matrices against {binary perturbations}.

The second type of parameter perturbations arises from variations in the weights of network links. These do not necessarily alter the zero-nonzero patterns of system matrices, {termed {\emph{structured perturbations}}}. They are typically modeled by additive perturbation matrices with specific zero/nonzero patterns, where nonzero entries indicate link weights subject to changes. Concepts like the structured controllability radius and structured stability radius quantify the smallest structured perturbations that cause a system to lose controllability and stability, respectively \cite{zhang2022real}. Although various algorithms address these issues
\cite{bianchin2016observability,zhang2022real}, they are often sub-optimal and rely on numerical algorithms prone to rounding errors, making them unsuitable for large-scale networks. 

{To demonstrate the distinct affections of the above two types of parameter perturbations on system properties, consider a linear system whose dynamic is described by
\begin{equation}\label{plant0} \begin{array}{c}
		\dot x(t)=Ax(t)+Bu(t),
\end{array} \end{equation}
where $A\in {\mathbb C}^{n\times n}$, $B\in {\mathbb C}^{n\times m}$, $x(t)$ and $u(t)$ are the state vector and input vector, respectively. Consider an example of $(A,B)$
 as
 $$[A, B]=\left[\begin{array}{ccc|cc}
	0&a_{12}&a_{13}&0&0\\
	0&a_{22}&a_{23}&0&b_{22}\\
	a_{31}&0&0&b_{31}&0
\end{array}\right],$$
where $a_{12},a_{13},...,b_{22}$ are free parameters. Let $[\Delta A, \Delta B]$ be the perturbation imposed on $[A,B]$, where only the $(1,2)$th entry of $\Delta A$ and the $(2,2)$th entry of $\Delta B$ are not fixed to zero. Denote the values of these entries by $\Delta a_{12}$ and $\Delta b_{22}$, respectively. The perturbed system $(A+\Delta A,B+\Delta B)$ then reads as
$$[A+\Delta A,B+\Delta B]= \left[\begin{array}{ccc|cc}
	0&a_{12}+\Delta a_{12}&a_{13}&0&0\\
	0&a_{22}&a_{23}&0&b_{22}+\Delta b_{22}\\
	a_{31}&0&0&b_{31}&0
\end{array}\right].$$
When $(\Delta A, \Delta B)$ belongs to a binary perturbation, meaning that the perturbed entries must be zero, i.e., $\Delta a_{12}=-a_{12}$, $\Delta b_{22}=-b_{22}$,
by \cite[Theo  1]{generic}, for almost all $(a_{12},a_{13},\cdots,b_{22})\in {\mathbb C}^7$, $(A+\Delta A,B+\Delta B)$ is controllable (i.e., the structured system associated with $(A+\Delta A,B+\Delta B)$ is structurally controllable\footnote{The structural controllability theory states that a pair $(A,B)$ is structurally controllable, if the generic rank of $[A,B]$ equals $n$ and an irreducibility condition for $[A,B]$ holds; see \cite[Theo  1]{generic} for details.}).
By contrast, if $(\Delta A, \Delta B)$ belongs to a structured perturbation, upon letting $\Delta b_{22}=-b_{22}$ and $\Delta a_{12}=(a_{22}a_{13})/a_{23}-a_{21}$ ($a_{23}\ne 0$), it is easy to verify that ${\rm rank}\, [A+\Delta A,B+\Delta B]\le 2$. According to the Popov-Belevitch-Hautus (PBH) test, for all $(a_{12},a_{13},\cdots,b_{22})\in {\mathbb C}^7$, in case $a_{23}\ne 0$, $(A+\Delta A, B+\Delta B)$ is uncontrollable.}

The above example motivates us to investigate the second type of perturbations within the structured system framework.
To the best of our knowledge, {except for \cite{full-version-tac} focusing on a special case (detailed subsequently), no prior work has considered such an issue}. In this paper, we address this gap by studying the {\emph{generic rank resilience (GRR)}} problem, {which examines to what extent the rank of realizations of a pattern matrix can be preserved generically against all structured perturbations with a given zero-nonzero pattern.} {Note that rank properties are crucial in applications to structured systems since most system-theoretic properties are characterized in terms of rank conditions \cite{belabbas2022structural,shali2021properties}.}
A pattern matrix is defined to be {\emph{structurally rank $r$ resilient against a perturbation pattern}} {if} almost all realizations of this pattern matrix maintain a rank of at least $r$ under arbitrary complex-valued realizations of the perturbation pattern.
{Unlike binary perturbations, these perturbations do not necessarily result in the addition/removal of free parameters, leading to parameter dependence in the perturbed matrices. This renders the conventional structured matrix theory inadequate, as it assumes independence among nonzero entries \cite{Murota_Book,generic}.}

The GRR problem is closely related to the low-rank matrix completion (LRMC) problem. LRMC involves completing a matrix with partially observed entries to achieve a desired low rank, with wide applications in recommendation systems \cite{bennett2007netflix}, image/video compression \cite{oseledets2011tensor}, reduced-order controller design, and system identification \cite{Fazel2002lowrank}, among others. Various techniques for LRMC have been developed \cite{candes2010power}. Recently, \cite{singer2010uniqueness,Kiraly-algebraic-2015,bernstein2020typical} have considered generic behaviors of LRMC using tools from the rigidity theory. These endeavors primarily focus on completing a {\emph{full}} matrix. However, in GRR, the completion matrices are structured (with partial entries fixed to zero), making {decomposition-based} and random {sampling} approaches in \cite{bernstein2020typical,Kiraly-algebraic-2015} unsuitable.\footnote{Specifically, considering $M(n,m,r)$ as the set of $n\times m$ complex full matrices with rank at most $r$, the decomposition-based method in \cite{bernstein2020typical,Kiraly-algebraic-2015} relies on the map {from ${\mathbb C}^{n\times r}\times {\mathbb C}^{r\times m}$ to $M(n,m,r)$: $(U,V)\to UV$. However, such a map does not exist when completions are structured since for some $(U,V)\in {\mathbb C}^{n\times r}\times {\mathbb C}^{r\times m}$, $UV$ may not have a required zero-nonzero pattern.}}

Towards the GRR problem, our contributions are summarized as follows. Firstly, we establish a generic property within GRR, that is, either almost all realizations of a pattern matrix are rank $r$ resilient against structured perturbations with a given zero-nonzero pattern, or almost all are not (Section \ref{sec-gene}). This highlights the dominance of structure in determining rank resilience. Our proof uses elementary tools from algebraic geometry and extends to the low-rank matrix pencil completion, {as in Section \ref{application}}. Secondly, we provide novel combinatorial conditions for a rectangular pattern matrix to lose full rank under structured perturbations (Section \ref{sec-k-1}). These conditions can be verified solely via computing some maximum bipartite matchings. 
Thirdly, we generalize these results to provide conditions for rectangular pattern matrices to lose any prescribed rank under structured perturbations (Section \ref{sec-k}). Finally,
as applications, we show how our results apply to the resilience analysis of controllability and input-state observability of structured descriptor systems, as well as characterizing zero structurally fixed modes.  (Section \ref{application}).   {Unlike \cite{full-version-tac}, which focuses on the controllability of single-input normal systems, our analysis of genericity in descriptor systems employs the low-rank matrix pencil completion,  not requiring the controllability matrix that is generically inaccessible for descriptor systems.  The problem addressed in \cite{full-version-tac} involves a specific case of the GRR problem: determining the condition for a pattern matrix of dimension \( n \times (n+1) \) to lose rank $1$ when only one entry is perturbed. In contrast, our work encompasses arbitrary rectangular pattern matrices and investigates the conditions under which they lose any given rank under arbitrary perturbation patterns.  Key to our results are a novel basis preservation principle and the dimension analysis of the involved algebraic variety.}

{\bf Notations}: Let ${\mathbb N}$ be the set of natural numbers and ${{\mathbb N}^+\doteq \{p\in {\mathbb N}: p\ge 1\}}$.  Given $p\in {\mathbb N}^+$, let ${\mathcal J}_p\doteq \{1,...,p\}$. For $n, m\in {\mathbb N}$ and $m\ge n$, define ${{\mathcal J}}^{n}_m \doteq \{{\mathcal I}\subseteq {\mathcal J}_m: |{\mathcal I}|=n\}$.  For a $p\times q$ matrix $M$, we use $M[{\mathcal I}_1,{\mathcal I}_2]$ to denote the submatrix of $M$ whose rows are indexed by ${\mathcal I}_1\subseteq {\mathcal J}_p$ and columns by ${\mathcal I}_2\subseteq {\mathcal J}_q$. In particular,  $M[{\mathcal J}_p,{\mathcal I}_2]$ will be denoted by $M[:,{\mathcal I}_2]$; similarly, $M[{\mathcal I}_1,{\mathcal J}_q]$ by $M[{\mathcal I}_1,:]$. {For a square matrix $M$, $\det(M)$ takes its determinant.} For a $p\times 1$ vector $a$, $a_{{\mathcal I}_1}$ stands for the sub-vector of $a$ indexed by ${\mathcal I}_1$, ${\mathcal I}_1\subseteq {\mathcal J}_p$. For two matrices $M_1,M_2$ with compatible dimensions, $[M_1;M_2]=[M_1^{\intercal},M_2^{\intercal}]^{\intercal}$. $0_n$ is the $n$-dimensional zero vector. {${\rm Im} (M)$ and  ${\rm null}(M)$ denote the column space and the null space of $M$, respectively.}

\vspace*{-0.5mm}
\section{Rank resilience of pattern matrices }\label{prob-formulation} 

In structured system theory \cite{generic}, a pattern matrix is a matrix whose entries are chosen from symbols $\{0,*\}$. Here, $0$ stands for a fixed zero entry and $*$ for an entry that can take arbitrary values independently. Let $\{0,*\}^{n\times m}$ be the set of all $n\times m$ pattern matrices. Given a pattern matrix ${\mathcal M}\in \{0,*\}^{n\times m}$, define a set
${\mathcal P}({\mathcal M})\doteq\left\{M\in {\mathbb C}^{n\times m}: M_{ij}=0 \ {\text{if}}\ {\mathcal M}_{ij}=0 \right\}$. Any $M\in {\mathcal {\mathcal P}({\mathcal M})}$ is called a realization of $\mathcal M$. {A {\emph{generic realization}} of ${\mathcal M}\in \{0,*\}^{n\times m}$ is a realization of ${\mathcal M}$ by assigning independent parameters to the $*$ entries,
i.e., parameters that do not satisfy any nontrivial polynomial equations \cite{Murota_Book}.}



\begin{definition}\label{generic-rank-def} \cite[p.38]{Murota_Book}
For an ${\mathcal M}\in \{0,*\}^{n_1\times n_2}$, its generic rank, denoted by ${\rm grank}({\mathcal M})$, is the rank it achieves for almost all its realizations, which also equals the maximum rank $\mathcal M$ achieves over all its realizations.
\end{definition}

%

To study the rank resilience of pattern matrices against structured perturbations, {we introduce a new type of pattern matrix, whose entries are specified by one of the three symbols $\{0,*,?\}$. The symbol $0$ stands for a fixed zero entry, $*$ for an entry that can take values independently, and $?$ for an {\emph{unspecified entry}}. An unspecified entry can take arbitrary values in the complex field $\mathbb C$, either independently of other entries or dependent on other entries (see Example \ref{comp-example}). The $*$ and $?$ entries are collectively called nonzero entries. Their distinctions will be further elaborated subsequently.}


Let $\{0,*,?\}^{n\times m}$ be the set of all $n\times m$ pattern matrices with entries from $\{0,*,?\}$, and likewise define $\{0,?\}^{n\times m}$. From a pattern matrix ${\mathcal M}\in \{0,*,?\}^{n\times m}$, we extract two pattern matrices ${\mathcal M}^o\in \{0,*\}^{n\times m}$ and ${\mathcal M}^{p}\in \{0,?\}^{n\times m}$, in which ${\mathcal M}^o_{ij}=*$ if ${\mathcal M}_{ij}=*$ and ${\mathcal M}^o_{ij}=0$ otherwise, and ${\mathcal M}^{p}_{ij}=?$ if ${\mathcal M}_{ij}=?$ and ${\mathcal M}^{p}_{ij}=0$ otherwise.  That is, ${\mathcal M}^o$ (resp., ${\mathcal M}^p$) is obtained from ${\mathcal M}$ by preserving only the $*$ entries (resp., $?$ entries) and putting $0$ to the remaining entries.
Here, ${\mathcal M}^o$ represents the pattern of the original unperturbed matrix and ${\mathcal M}^p$ a perturbation pattern. Let ${\mathcal P}({\mathcal M}^p)=\left\{M\in {\mathbb C}^{n\times m}: M_{ij}=0 \ {\text{if}}\ {\mathcal M}^p_{ij}=0 \right\}$ be the set of structured perturbations specified by ${\mathcal M}^p$. {Upon letting $n_*$ be the number of $*$ entries in ${\mathcal M}^o$, it follows that ${\mathcal P}({\mathcal M}^{o})$ corresponds to the vector space ${\mathbb C}^{n_*}$.} {The {\emph{$k$-generic rank resilience problem}}, abbreviated as $k$-GRR, is posed as follows.}

{\emph{{\bf $k$-GRR}: Given ${\mathcal M}\in \{0,*,?\}^{n\times m}$ and $k\in {\mathbb N}$, verify whether for {almost all} $M\in {\mathcal P}({\mathcal M}^o)$ (i.e., all except for some proper variety in the parameter space corresponding to ${\mathcal M}^o$), there is a structured perturbation $\Delta M\in {\mathcal P}({\mathcal M}^p)$ such that ${\text {rank}} (M+\Delta M)\le n-k$.}}


The $k$-GRR, as its name suggests, asks whether a generic realization of ${\mathcal M}^o$ can be perturbed to possess a low rank at most $n-k$ by some structured perturbation in ${\mathcal P}({\mathcal M}^p)$.  Consequently, the $*$ entries in ${\mathcal M}$ are referred to as \textit{generic entries}, indicating that they assume generic values in the context of $k$-GRR, i.e., values from a full-dimensional space excluding certain hypersurfaces.  Two related definitions are introduced as follows.


{\begin{definition} \label{def-rank-resilience}
Given a  matrix $M\in {\mathcal P}({\mathcal M}^o)$ and $r\in {\mathbb N}$, $M$ is called {\emph{rank $r+1$ resilient against a perturbation pattern ${\mathcal M}^p$}}, if ${\text {rank}}  (M+\Delta M)\ge r+1$ for any $\Delta M\in {\mathcal P}({\mathcal M}^p)$. Conversely, if there is a $\Delta M\in {\mathcal P}({\mathcal M}^p)$ such that ${\text {rank}}  (M+\Delta M)\le r$, $M$ is called {\emph{rank $r$ completable against ${\mathcal M}^p$}}.
\end{definition}

\begin{definition} \label{def_structural_rank_resilience}
Given ${\mathcal M}\in \{0,*,?\}^{n\times m}$ and $k\in {\mathbb N}$, if for almost all $M\in {\mathcal P}({\mathcal M}^o)$, $M$ is {\emph{rank $r+1$ resilient against ${\mathcal M}^p$}}, ${\mathcal M}^o$ is called {\emph{structurally rank $r+1$ resilient against ${\mathcal M}^p$}}.  If for almost all $M\in {\mathcal P}({\mathcal M}^o)$, $M$ is {\emph{rank $r$ completable against ${\mathcal M}^p$}}, ${\mathcal M}^o$ is called {\emph{structurally rank $r$ completable against ${\mathcal M}^p$}}.    \end{definition}}

In Definition \ref{def-rank-resilience}, the terminology ``rank $r$ completability'' comes from the well-known LRMC problem, {which can be formulated as the problem of completing the missing entries of a matrix with partially observed entries to make it have a desired low rank \cite{Fazel2002lowrank,candes2010power}. Here, the perturbed entries ($?$ entries) prescribed by ${\mathcal M}^p$ can be regarded as missing entries to be completed and nonzero entries ($*$ entries) in ${\mathcal M}^o$ as observed entries, while the remaining fixed zero entries indicate the known sparsity pattern.}
In this way, when ${\mathcal M}^o$ is structurally rank $r\doteq n-k$ completable against ${\mathcal M}^p$, we interchangeably say that {\emph{the generic low-rank matrix associated with $({\mathcal M}, k)$, abbreviated as GLRM$({\mathcal M},k)$, is completable}}, otherwise GLRM$({\mathcal M},k)$ is incompletable. Notice that $({\mathcal M},k)$ encodes all information in the addressed $k$-GRR. In the next section, we shall show the rank $r+1$ ($\forall r\in {\mathbb N}$) resilience is a generic property {in the sense that there cannot be a third case, i.e., for any ${\mathcal M}$, there cannot be a set ${\mathcal P}\subseteq {\mathcal P}({\mathcal M}^o)$,  with which both ${\mathcal P}$ and ${\mathcal P}({\mathcal M}^o)\backslash {\mathcal P}$ have nonzero measure in ${\mathcal P}({\mathcal M}^o)$, such that each $M\in {\mathcal P}$ is rank $r$ completable against ${\mathcal M}^p$, while every $M'\in {\mathcal P}({\mathcal M}^o)\backslash {\mathcal P}$ is rank $r+1$ resilient against ${\mathcal M}^p$, exhibiting a law of excluded middle.} When considering the $k$-GRR of a pattern matrix ${\mathcal M}^o$ against ${\mathcal M}^p$, if certain $*$ entries in ${\mathcal M}^o$ and $?$ entries in ${\mathcal M}^p$ are of the same positions, in constructing GLRM$({\mathcal M},k)$, it suffices to put $?$ to these entries in ${\mathcal M}$.

Throughout this paper, given $\mathcal{M}\in\{0,*,?\}^{n\times m}$,  let $p_*\doteq(p_{*1},...,p_{*n_*})$ and $p_{?}\doteq (p_{?1},...,p_{?n_?})$ respectively be the vectors of parameters for the $*$ entries and $?$ entries in ${\mathcal M}$, where $n_*$ ($n_?$) is the number of $*$ entries ($?$ entries) in ${\mathcal M}$. Moreover, let $X(p_*,p_?)$ ($M(p^*)$) be a realization by substituting the parameters $p^*$ and $p_?$ into ${\mathcal M}$ (${\mathcal M}^{o}$). Borrowing the terminology from LRMC, we occasionally call $X(p_*,p_?)$ {\emph{a completion}} of ${\mathcal M}$.
In the following, a comprehensible example is given to illustrate the $k$-GRR.

\begin{example}\label{comp-example} Consider ${\mathcal M}=\left[
	\begin{array}{cccc}
		* & ? & * & ?\\
		* & * & ? & 0\\
	\end{array}
	\right]
	$. Then ${\mathcal M}^o=\left[
	\begin{array}{cccc}
		* & 0 & * & 0\\
		* & * & 0 & 0\\
	\end{array}
	\right]$ and ${\mathcal M}^p=\left[
	\begin{array}{cccc}
		0 & ? & 0 & ?\\
		0 & 0 & ? & 0\\
	\end{array}
	\right]$. Consequently,
	${\mathcal P}({\mathcal M}^o)=\left\{\left[
	\begin{array}{cccc}
		a & 0 & b & 0\\
		c & d & 0 & 0\\
	\end{array}
	\right], (a,b,c,d)\in {\mathbb C}^4\right\}
	$ and 	${\mathcal P}({\mathcal M}^p)=\left\{\left[
	\begin{array}{cccc}
		0 & e & 0 & g\\
		0 & 0 & f & 0\\
	\end{array}
	\right], (e,f,g)\in {\mathbb C}^3\right\}.
	$  
	If $k=1$, by making all $2\times 2$ minors of $M+\Delta M$ equal zero with $M\in {\mathcal P}({\mathcal M}^o)$ and $\Delta M\in {\mathcal P}({\mathcal M}^p)$, it turns out that for all $(a,b,c,d)\in {\mathbb C}^4$ except the set $\{(a,b,c,d)\in {\mathbb C}^4: c=0,ad\ne 0 \ {\text {or}}\ a=0, bc\ne 0 \}$, there exists a set of complex values for the $?$ entries making ${\rm rank}(M+\Delta M)\le 1$ {(e.g., $e=(ad)/c,f=(bc)/a$, and $g=0$ when $ac\ne 0$)}. Therefore, ${\mathcal M}^o$ is structurally rank $1$ completable against ${\mathcal M}^p$. On the other hand, if we change $\mathcal M$ to ${\mathcal M}'=\left[
	\begin{array}{cccc}
		* & ? & * & ?\\
		0 & * & ? & 0\\
	\end{array}
	\right]
	,$ then a matrix $M'\in {\mathcal P}({\mathcal M}^{o'})$ can be expressed as $\left[
	\begin{array}{cccc}
		a & 0 & b & 0 \\
		0 & d & 0 & 0 \\
	\end{array}
	\right]
	.$ For each $(a,b,d)\in {\mathbb C}^3$ except the set $\{(a,b,d)\in {\mathbb C}^3: ad=0\}$, there is no perturbation $\Delta M'\in {\mathcal P}({\mathcal M}^{'p})$ making the corresponding $M'+\Delta M'$ have rank less than $2$. Therefore, ${\mathcal M}^{o'}$ is structurally rank $2$ resilient against ${\mathcal M}^{'p}$.
\end{example}


Without losing generality and to avoid trivial cases, the following assumption is made throughout this paper.

\begin{assumption}\label{throughout_assumption} In GLRM$({\mathcal M}, k)$ with ${\mathcal M}\in \{0,*,?\}^{n\times m}$,
	$m\ge n$ and ${\rm grank}({{\mathcal M}^{o}})> n-k$.
\end{assumption}

If $m<n$, we can take the transpose. If ${\rm grank}({{\mathcal M}^{o}})\le n-k$, then GLRM$({\mathcal M},k)$ is immediately completable by letting all $?$ entries be zero. It is known that the generic rank of ${\mathcal M}^{o}$ can be determined by computing a maximum bipartite matching (see \cite[p.39]{Murota_Book}), e.g., via the Hopcroft-Karp algorithm. This means Assumption \ref{throughout_assumption} can be efficiently verified.

\section{Genericity in $k$-GRR} \label{sec-gene}

{In this section, we justify and establish the underlying generic properties { of $k$-GRR}. Our proof is also easily extended to show the genericity of resilience of several properties in structured (descriptor) systems as shown in Section \ref{application}. 

	To proceed with our proof, the following definitions and certifications in algebraic geometry are needed (see \cite{Hartshorne2013}). Let
	$$f_1(x_1,...,x_n)=0,\cdots, f_t(x_1,...,x_n)=0 $$
	be a system of polynomial equations, denoted by $(f_1,...,f_t)$, where $f_i$ ($i=1,...,t$) are polynomials in the variables $(x_1,...,x_n)$ with real/complex coefficients. A system of equations $(f_1,...,f_t)$ is said to be inconsistent if there is no solution for $(x_1,...,x_n)\in {\mathbb C}^n$, and is underdetermined if it has fewer equations than unknowns, i.e., $t<n$.

	\begin{lemma}(Hilbert's Nullstellensatz and dimension of solutions for undetermined systems) \cite{Sombra1999},\cite[Exer 1.9]{Hartshorne2013}  \label{system_polynomial} The following statements are true with respect to (w.r.t.) the solutions of a system of polynomial equations $(f_1,...,f_t)$ in variables $(x_1,...,x_n)$:
		
		1) $(f_1,...,f_t)$ is inconsistent if and only if there exist {polynomials $p_i(x_1,...,x_t)$ in the variables $(x_1,...,x_t)$} {(denoted by $p_i$ subsequently for notational simplicity)}, $i=1,...,t$, such that \begin{equation}\label{nulltheorem}\sum \nolimits_{i=1}^t {p_i(x_1,...,x_t)}f_i(x_1,...,x_n)=1.\end{equation}
		
		2) If $(f_1,...,f_t)$ is underdetermined, then it has either infinitely many complex solutions or is inconsistent.  Moreover, if $(f_1,...,f_t)$ has solutions, then the set of all complex solutions forms an algebraic set of dimension at least $n - t$.\footnote{\label{algebraic_dimension_foot1}This means there are at most $t$ variables that are algebraic over the remaining variables which form a transcendence basis of the corresponding function field.  The dimension of an algebraic set roughly refers to the number of independent parameters required to describe it; see \cite[Chap 8]{Milne2003}.}
	\end{lemma}

	\begin{lemma}[Effective Nullstellensatz] \cite{Sombra1999}  \label{effective_nullste} Let $f_1,...,f_t$ be $t$ polynomials in $n\ge 1$ variables of total degree $d_1\ge \cdots \ge d_t$ (the total degree of a multivariate polynomial $f$, given by $\deg(f)$, is the maximum degree of a monomial in $f$). If there exist polynomials $p_i$ ($i=1,...,t$) such that $\sum \nolimits_{i=1}^t p_if_i=1$, then they can be chosen such that $\deg(p_if_i)\le 2d_t \prod \nolimits_{j=1}^{\min(n,t)-1} d_j$.\footnote{If $n=1$, this inequality collapses to $\deg(p_if_i)\le2d_1$.}\end{lemma}



	\begin{lemma} \label{rank_diff_generic} Consider a $p\times q$ matrix $M$ whose entries are polynomials of variables $(x_1,...,x_n)$ with real or complex coefficients ($p,q\ge 1$). Then, the value of the function $r(x_1,...,x_n)={\rm rank}(M) - {\rm rank}(M[{\mathcal J}_p\backslash \{p\},:])$ is generic in ${\mathbb C}^n$ in the sense that for all $(x_1,...,x_n)\in {\mathbb C}^n$ except a set of measure zero, $r(x_1,...,x_n)$ takes the same value ($0$ or $1$).
	\end{lemma}
	
	\begin{proof}  Note that the rank of a matrix with polynomial entries is generic. This is because the rank equals the maximum dimension of a nonsingular square submatrix.  The zero set of its determinant forms a proper algebraic variety in the corresponding parameter space \cite[Chap.2.1]{Murota_Book}. Therefore, for all $(x_1,...,x_n)\in {\mathbb C}^n\backslash {\mathcal P}_1$, where ${\mathcal P}_1$ denotes some proper algebraic variety in ${\mathbb C}^n$, we have
		${\rm rank}(M(x_1,...,x_n))= {\rm grank}(M),$
		where $M(x_1,...,x_n)$ denotes the matrix obtained by substituting the values $(x_1,...,x_n)$.
		Similarly, for all  $(x_1,...,x_n)\in {\mathbb C}^n\backslash {\mathcal P}_2$, where ${\mathcal P}_2$ denotes some proper algebraic variety in ${\mathbb C}^n$, we have
		$ {\rm rank} (M(x_1,...,x_n)[{\mathcal J}_p\backslash \{p\},:])= {\rm grank}(M[{\mathcal J}_p\backslash \{p\},:]).$
		Therefore, for all $(x_1,...,x_n)\in {\mathbb C}^n\backslash ({\mathcal P}_1\cup {\mathcal P}_2)$,
		$r(x_1,...,x_n)= {\rm grank} (M) - {\rm grank}(M[{\mathcal J}_p\backslash \{p\},:]),$
		which value is generic in ${\mathbb C}^n$ since ${\mathcal P}_1\cup {\mathcal P}_2$ is still a proper algebraic variety in ${\mathbb C}^n$.
	\end{proof}
	
	\begin{theorem} \label{feasibility_theorem} Given $({\mathcal M},k)$,   either almost all $M\in {\mathcal P}({\mathcal M}^{p})$ are rank $n-k$ completable against ${\mathcal M}^p$, or almost all $M\in {\mathcal P}({\mathcal M}^{p})$ are rank $n-k+1$ resilient against ${\mathcal M}^p$.
	\end{theorem}
	
	\begin{proof}  Let $X(p_*,p_?)$ be a completion of $\mathcal M$ with parameters $p_*$ and $p_?$ defined in Section \ref{prob-formulation}. To make ${\rm rank}(X(p_*,p_?)) \le  n- k$, it is necessary and sufficient that every $(n-k+1)\times (n-k+1)$ minor of $X(p_*,p_?)$ is zero, that is,
		\begin{equation}\label{non_feasible_constraint}\begin{array}{c} \det X(p_*,p_?)[{\mathcal I}_1, {\mathcal I}_2]=0, \ \forall \  {\mathcal I}_1 \in {\mathcal J}_n^{n-k+1},  {\mathcal I}_2 \in {\mathcal J}_m^{n-k+1}.\end{array} \end{equation}
		The above constraint induces a system of polynomial equations in the variables $p_?$ with coefficients polynomials of $p_*$.
		If there exists a set of polynomials $\{f_{{\mathcal I}_1,{\mathcal I}_2}\}_{{\mathcal I}_1 \in {\mathcal J}_n^{n-k+1}, {\mathcal I}_2 \in {\mathcal J}_m^{n-k+1}}$ in the variables $p_?$, such that
		\begin{equation}\label{feasible_condition} \sum \limits_{{{\mathcal I}_1 \in {\mathcal J}_n^{n-k+1}, {\mathcal I}_2 \in {\mathcal J}_m^{n-k+1}}} f_{{\mathcal I}_1,{\mathcal I}_2} \det X(p_*,p_?)[{\mathcal I}_1, {\mathcal I}_2]=1,\end{equation}
		then, from Lemma \ref{effective_nullste}, they can be chosen such that
	$\deg(f_{{\mathcal I}_1,{\mathcal I}_2}\det X(p_*,p_?)[{\mathcal I}_1, {\mathcal I}_2])\le 2(n-k+1)^{n_?},$
		noting that the total degree in $\det X(p_*,p_?)[{\mathcal I}_1, {\mathcal I}_2]$ is at most $n-k+1$. Let $r_{{\mathcal I}_1,{\mathcal I}_2}\le 2(n-k+1)^{n_?}$ be an upper bound of $\deg(f_{{\mathcal I}_1,{\mathcal I}_2})$. For each pair $({\mathcal I}_1,{\mathcal I}_2)\in {\mathcal J}_n^{n-k+1}\times {\mathcal J}_m^{n-k+1}$, write $f_{{\mathcal I}_1,{\mathcal I}_2}$ as
		\begin{equation}\label{expression_coefficient}f_{{\mathcal I}_1,{\mathcal I}_2}=\sum \limits_{0\le r_1+\cdots+r_{n_?}\le r_{{\mathcal I}_1,{\mathcal I}_2}} a^{{{\mathcal I}_1,{\mathcal I}_2}}_{r_1...r_{n?}}p_{?1}^{r_1} \cdot...\cdot p^{r_{n_?}}_{?n_?}, \end{equation}where $\{a^{{{\mathcal I}_1,{\mathcal I}_2}}_{r_1...r_{n?}}\}$ are the unknown coefficients to be determined. Upon substituting (\ref{expression_coefficient}) into (\ref{feasible_condition}), and letting the coefficient for each non-constant monomial $\prod \nolimits_{i=1}^{n_?} p_{?i}^{r_i}$ be zero,  (\ref{feasible_condition}) is reduced to a finite system of linear equations of the following form
		\begin{equation} \label{linear_system}
			Q(p_*){\bf a}=[0_{N-1}^{\intercal},1]^{\intercal},
		\end{equation}
		where $\bf a$ is the vector obtained by stacking all the coefficients $\{a^{{{\mathcal I}_1,{\mathcal I}_2}}_{r_1...r_{n?}}\}$, $Q(p_*)$ is a matrix whose entries are polynomials of $p_*$, and $N-1$ is the total number of non-constant monomials $\prod \nolimits_{i=1}^{n_?} p_{?i}^{r_i}$. Eq. (\ref{linear_system}) has a solution for $\bf a$, if and only if
	${\rm rank}(Q(p_*))={\rm rank}([Q(p_*), {\bf e}]),$ with ${\bf e}=[0_{N-1}^{\intercal},1]^{\intercal}$, which is equivalent to
		\begin{equation}\label{finalcond} {\rm rank}(Q(p_*))-  {\rm rank}(Q(p_*)[{\mathcal J}_N\backslash \{N\},:])=1.\end{equation}
		Recall it has been proven in Lemma \ref{rank_diff_generic} that the value of the left-hand side of (\ref{finalcond}) is generic in ${\mathbb C}^{n_*}$. This means for almost all $p_*\in {\mathbb C}^{n_*}$, either (\ref{finalcond}) is satisfied and thus (\ref{feasible_condition}) has a solution for $\{f_{{\mathcal I}_1,{\mathcal I}_2}\}_{{\mathcal I}_1 \in {\mathcal J}_n^{n-k+1}, {\mathcal I}_2 \in {\mathcal J}_m^{n-k+1}}$, indicting that (\ref{non_feasible_constraint}) has no solution for $p_?\in {\mathbb C}^{n_?}$, or the contrary.
	\end{proof}

	Theorem \ref{feasibility_theorem} makes clear that it is the combinatorial structure of ${\mathcal M}$ that dominates the rank $r$ resilience of realizations of ${\mathcal M}^{o}$ against ${\mathcal M}^p$. This justifies the definition of $k$-GRR.

\begin{remark}\label{gene-real} {Note that the value of the function $r(x_1,...,x_n)$ in Lemma \ref{rank_diff_generic} is still generic in the real vector space ${\mathbb R}^n$.
Following the above proof, the generic property in Theorem \ref{feasibility_theorem} remains true even when the generic entries can only take real values (but the unspecified entries need to take complex values). However, when the unspecified entries are restricted to be real, the Hilbert's Nullstellensatz does not hold. Since the existence of real solutions of polynomials often depends on the non-negativeness of certain discriminants \cite{Hartshorne2013}, there may be a semi-algebraic set\footnote{A semi-algebraic set of ${\mathbb R}^{n_*}$ is a subset of ${\mathbb R}^{n_*}$ defined by a finite number of polynomial equations and inequalities.}of nonzero measure in the real vector space ${\mathbb R}^{n_*}$ such that the corresponding realizations are rank $r$ resilient against all real perturbations in ${\mathcal P}({\mathcal M}^p)$; meanwhile, the rest realizations, also corresponding to a semi-algebraic set of nonzero measure, are rank $r$ completable; see the following illustrative example. Nevertheless, the rank resilience against complex perturbations implies the same property against real perturbations. That is, conditions for the former property are sufficient for the latter.} \end{remark}

{\begin{example}
	Consider a pattern matrix ${\mathcal M}$ with its completion given by
{\footnotesize 	$${\mathcal M}=\left[ {\begin{array}{*{20}{c}}
		?&*&0&*\\
		0&?&*&*\\
		*&*&?&*\\
		*&*&*&?
		\end{array}} \right], X = \left[ {\begin{array}{*{20}{c}}
		{{a_{11}}}&{{a_{12}}}&0&{{a_{14}}}\\
		0&a_{22}&{{a_{23}}}&{{a_{24}}}\\
		{{a_{31}}}&{{a_{32}}}&a_{33}&{{a_{34}}}\\
		{{a_{41}}}&{{a_{42}}}&{{a_{43}}}&{{a_{44}}}
		\end{array}} \right].$$}Vanishing the $3\times3$ minor of the matrix
	obtained from $X$ by removing the second row and third column, and the matrix obtained by removing the third row and second column gives two expressions of $a_{44}$. The difference of them must vanish, yielding a quadratic equation in $a_{11}$ with coefficients being polynomials of the remaining parameters except for $a_{44}$ and $a_{11}$, given as follows:
$$\begin{array}{c} (a_{23}a_{34}a_{42} - a_{24}a_{32}a_{43})a_{11}^2 + [a_{12}a_{24}a_{31}a_{43} \\- a_{23}(a_{12}a_{34}a_{41} + a_{14}a_{31}a_{42} - a_{14}a_{32}a_{41}) - a_{14}a_{23}a_{32}a_{41}]a_{11}\\ + a_{12}a_{14}a_{23}a_{31}a_{41} = 0. \end{array}$$
The discriminant $\Delta$ of this equation w.r.t. $a_{11}$ is given by $$\begin{array}{c}\Delta=(a_{12}a_{23}a_{34}a_{41} - a_{12}a_{24}a_{31}a_{43} + a_{14}a_{23}a_{31}a_{42})^2 \\- 4a_{12}a_{14}a_{23}a_{31}a_{41}(a_{23}a_{34}a_{42} - a_{24}a_{32}a_{43})\end{array}.$$Hence, if $\Delta<0$, no real completions with rank $2$ exist. In addition, it can be verified via the symbolic computation tools in Matlab that if $\Delta\ge 0$, there exist real completions with rank $2$ in general. It turns out that both $\Delta<0$ and $\Delta\ge 0$ induce two semi-algebraic sets with nonzero measure.
\end{example}}

	\section{Conditions for completability of GLRM$({\mathcal M},1)$}  \label{sec-k-1}
	In this section, we give a necessary and sufficient combinatorial condition for the $k$-GRR with $k=1$.  Our approach is based on a key finding of this paper, namely, the basis preservation principle, expanded as follows. 
	
	\subsection{Basis preservation approach} \label{approach}
	Some global definitions are introduced first. Given an ${\mathcal M}\in \{0,*,?\}^{n\times m}$, for $i=1,...,m$, define ${\mathcal N}_{*i}=\{j\in {\mathcal J}_n: {\mathcal M}_{ji}= * \}$ as the set of row indices for the $*$ entries in the $i$th column of $\mathcal M$. Let ${\mathcal N}_{?i}$ be defined similarly for the $?$ entries, and ${\mathcal N}_i\doteq {\mathcal N}_{?i}\cup {\mathcal N}_{*i}$.
	For ${\mathcal I}\subseteq {\mathcal J}_m$, define
	${\mathcal M}({{\mathcal I}})\in \{0,*,?\}^{n\times m}$ as
	\begin{equation}\label{basis-def} [{\mathcal M}({\mathcal I})]_{ij}=
		\begin{cases} * & \text{if}\ j\in {\mathcal I}, \ {\mathcal M}_{ij}= ?\\
			{\mathcal M}_{ij} & {\text{otherwise,}}
	\end{cases}\end{equation}
	i.e., ${\mathcal M}({{\mathcal I}})$ is obtained from $\mathcal M$ by changing the $?$ entries in the columns indexed by ${\mathcal I}$ to $*$.
	For each ${\mathcal{M}}\in\{0,*,?\}^{n\times m}$, define $\hat {\mathcal M}\doteq {\mathcal M}({\mathcal J}_m)$, i.e., obtained by replacing all $?$
	entries of ${\mathcal M}$ with $*$ entries.
	
%
		
		\begin{definition}[Preservable basis] Given an ${\mathcal M}\in \{0,*,?\}^{n\times m}$, we call ${\mathcal I}\in {\mathcal J}^{n-1}_m$ a preservable basis of $\mathcal M$, if ${\rm grank}(\hat {\mathcal M}[:,{\mathcal I}])=n-1$.
		\end{definition}
		
		\begin{theorem}[Basis preservation principle]  \label{BasisPreservation}
			Under Assumption \ref{throughout_assumption}, GLRM$({\mathcal M},1)$ is completable, if and only if there exists a preservable basis ${\mathcal I}$ of $\mathcal M$, such that GLRM$({\mathcal M}({\mathcal I}),1)$ is completable.
		\end{theorem}
		
		Due to its lengthiness,	the proof is postponed to Section \ref{proof}. Theorem \ref{BasisPreservation} indicates that if GLRM$({\mathcal M},1)$ is completable, {\emph{we can always find a subset of its columns so that by replacing all unspecified entries therein with generic entries, the obtained columns can serve as a {\bf basis} of the space spanned by the corresponding completion of rank $n-1$}. As such, we call Theorem \ref{BasisPreservation} the basis preservation principle. This principle is promising in verifying the completability of GLRM$({\mathcal M},1)$, transforming it into an equivalent problem with $({\mathcal M}({\mathcal I}),1)$. The latter problem can be relatively easily solved via the basis preservation approach shown below.

			\begin{proposition}\label{BasisApproach} Under Assumption \ref{throughout_assumption}, suppose that there exists a set of columns of $\mathcal M$ indexed by ${\mathcal I}\in {\mathcal J}^{n-1}_m$ such that ${\mathcal M}[:, {\mathcal I}]$ contains no $?$ entries and ${\rm grank } ({\mathcal M}[:, {\mathcal I}])=n-1$. Moreover, define ${\mathcal B}=\{{\mathcal K}\in {\mathcal J}^{n-1}_n: {\rm grank} ({\mathcal M}[{\mathcal K}, {\mathcal I}])=n-1\}$, and let ${\mathcal B}^*= \{{\mathcal J}_n\backslash \Omega: \Omega\in {\mathcal B}\}$. GLRM$({\mathcal M},1)$ is completable, if and only if for each $i\in {\mathcal J}_{m}\backslash {\mathcal I}$, one of the following conditions holds:
				\begin{itemize}
					\item[1)] ${\mathcal B}^*\cap {\mathcal N}_{*i}=\emptyset$;
					\item[2)] If ${\mathcal B}^*\cap {\mathcal N}_{*i}\neq \emptyset$, then ${\mathcal B}^*\cap {\mathcal N}_{?i}\neq \emptyset$.
				\end{itemize}				
			\end{proposition}
			
			The proof is given in Section \ref{proof}. Combining Theorem \ref{BasisPreservation} and Proposition \ref{BasisApproach} yields a complete solution for the completability of GLRM$({\mathcal M},1)$. The proof is omitted due to its obviousness.
			
			\begin{theorem} \label{SolutionKone} Under Assumption \ref{throughout_assumption}, GLRM$({\mathcal M},1)$ is completable, if and only if there exists a preservable basis $\mathcal I$ of $\mathcal M$, such that for each $i\in {\mathcal J}_m\backslash {\mathcal I}$, either ${\mathcal B}^*\cap {\mathcal N}_{*i}=\emptyset$, or ${\mathcal B}^*\cap {\mathcal N}_{?i}\neq \emptyset$,
				where ${\mathcal B}= \{{\mathcal K}\in {\mathcal J}^{n-1}_n: {\rm grank}({\mathcal M}({\mathcal I})[{\mathcal K}, {\mathcal I}])=n-1\}$, ${\mathcal B}^*= \{{\mathcal J}_n\backslash \Omega: \Omega\in {\mathcal B}\}$.

			\end{theorem}
			

			\begin{example} \label{exp3} Consider the GLRM$({\mathcal M},1)$ with {\footnotesize $${\mathcal M}=\left[
				\begin{array}{cccc}
					* & * & 0 & ? \\
					0 & ? & * & 0 \\
					* & * & ? & * \\
				\end{array}
				\right].
				$$}The corresponding sets are ${\mathcal N}_{*1}=\{1,3\}$, ${\mathcal N}_{*2}=\{1,3\}$, ${\mathcal N}_{*3}=\{2\}$, ${\mathcal N}_{*4}=\{3\}$, ${\mathcal N}_{?1}=\emptyset$, ${\mathcal N}_{?2}=\{2\}$, ${\mathcal N}_{?3}=\{3\}$, and ${\mathcal N}_{?4}=\{1\}$. Any subset of $\{1,...,4\}$ containing two elements can be a preservable basis. It is found that for the preservable basis $\{1,2\}$ (or $\{1,3\}$, $\{1,4\}$), conditions in Theorem \ref{SolutionKone} are satisfied, while not for other preservable bases. This indicates that GLRM$({\mathcal M},1)$ is completable. 
			\end{example}
			
{\begin{remark}By the fundamental theorem of algebra and the definition of determinant, when $m=n$, GLRM$({\mathcal M},1)$ is completable if and only if there exist $n$ nonzero entries in ${\mathcal M}$, one of which is a $?$ entry, such that these entries are in different rows and columns. This condition is equivalent to the statement in Theorem \ref{SolutionKone} for a square matrix ${\mathcal M}$. To see this, assume there is a $?$ entry in the $i$th row and $j$th column of $\mathcal M$ and this entry is among the aforementioned $n$ nonzero entries. Then, ${\mathcal I}\doteq {\mathcal J}_n\backslash \{j\}$ is a preservable basis. From ${\rm grank}({\mathcal M}({\mathcal I})[{\mathcal J}_n\backslash \{i\},{\mathcal I}])=n-1$, it follows that $i\in {\mathcal B}^*$.  Thus, $i\in {\mathcal B}^*\cap {\mathcal N}_{?j}\ne \emptyset$, i.e., the condition in Theorem \ref{SolutionKone} is satisfied.  The reverse direction follows similarly. A graph-theoretic representation of Theorem \ref{SolutionKone} can be found in \cite{arxiv-version}. \end{remark}}			
			
			The promising feature of Theorem \ref{SolutionKone} is that it only involves some basic combinatorial operations, such as generic rank computations and the set union/intersection operations. Concerning the computational complexity of Theorem \ref{SolutionKone}, testing whether $n-1$ columns of $\mathcal M$ form a preservable basis incurs at most $O(n^2\sqrt{n})$ complexity using the Hopcroft-Karp algorithm \cite{Murota_Book}. In the worst case, it requires testing $\binom{m}{n-1}$ preservable bases. For a preservable basis, determining ${\mathcal B}^*$ incurs at most $O(n\times n^2\sqrt{n})$ complexity using the Hopcroft-Karp algorithm, and checking the two conditions in Theorem \ref{SolutionKone} costs complexity $O(mn)$. {To sum up, the worst-case computational complexity of Theorem \ref{SolutionKone} is $O(m^{\min\{n-1,m-n+1\}}(n^{3.5}+mn))$, since $\binom{m}{n-1}=\binom{m}{m-n+1}\le \min\{m^{n-1},m^{m-n+1}\}$.} When $m-n$ is bounded by some constant (e.g., ${\mathcal M}$ is a square matrice or ${\mathcal M}$ is $[A,B]$ of a single-input system in (\ref{plant0})), this achieves a polynomial time complexity in $n$ and $m$. However, without that constraint, the complexity may increase exponentially with $n$ and $m$.
			In practice, we can test a limited number of randomly selected preservable bases rather than enumerate all the possible ones. Following this idea, a randomized implement of Theorem \ref{SolutionKone} is given in Algorithm \ref{alg1-theo4}. In this algorithm, $T_m$ is the maximum number of samples of $(n-1)$-columns. The larger $T_m$ is, the higher the probability will be that the returned answer is right, with the price of an increased computational burden. More precisely, if the GLRM(${\mathcal M},1$) is incompletable, Algorithm \ref{alg1-theo4} always returns the right answer; otherwise, Algorithm \ref{alg1-theo4} returns the right answer with probability $P_r\ge 1- (1-P_0)^{T_m}$ with $P_0=1/\binom{m}{n-1}$. {Since there are at most $T_m$ samples of $(n-1)$-columns, following the previous analysis, the computational complexity of Algorithm \ref{alg1-theo4} is $O(T_m (mn+n^{3.5}))$.}

		{\small{	\begin{algorithm} 
				{{{{
								\caption{: Practical verification of Theorem \ref{SolutionKone} on GLRM$({\mathcal M},1)$ with ${\mathcal M}\in \{0,*,?\}^{n\times m}$ via random-sampling} 
								\label{alg1-theo4} 
								\begin{algorithmic}[1] 
									\State Initialize $flag=0$, $ite=0$;
									\While{$flag=0$ and $ite<= T_m$}
									\State ${\mathcal I}={\rm randperm}(m,n-1)$; {$^{1}$}
									\State $ite=ite+1$;
									\If{${\mathcal I}$ is a preservable basis of $\mathcal M$}
									\State Determine ${\mathcal B}^*$;
									\If{ Condition 1) or 2) in Theorem \ref{SolutionKone} is satisfied }
									\State $flag=1$;
									\State Break;
									\EndIf
									\EndIf
									\EndWhile
									\State {Return $flag$ (if $flag=1$, the answer is completable; otherwise terminate with being incompletable {\emph{regardless of the actual answer}})}.
						\end{algorithmic}}}
				}}
				{\footnotemark[1]{\emph{${\rm randperm}(m,n-1)$  returns a set containing $n-1$ distinct integers
							selected randomly from $\{1,...,m\}$. }}}
			\end{algorithm}
		}}
			
			\subsection{Graph-theoretic form} \label{sec-graph}
			In this subsection, we give an equivalent bipartite graph form of Theorem \ref{SolutionKone}. 
			
			Some basic definitions in graph theory are first introduced \cite{Murota_Book}. A bipartite graph is a graph whose vertices can be partitioned into two parts, such that there is no edge within each part. A matching is a set of edges such that any two of them do not share a common vertex. A maximum matching of a graph $\mathbb G$ is the matching with the largest number of edges, whose cardinality is denoted by ${\rm mt}({\mathbb G})$. 
			
			Given ${\mathcal M}\in \{0,*,?\}^{n\times m}$, let ${\mathbb G}({\mathcal M})=({\mathbb V}^+,{\mathbb V}^-, {\mathbb E},{\mathcal M})$ be the bipartite graph associated with $\mathcal M$, with vertex sets ${\mathbb V}^{+}=\{v^+_1,...,v^+_n\}$, ${\mathbb V}^{-}=\{v^-_1,...,v^-_m\}$, and edge set ${\mathbb E}=\{(v^+_i,v^-_j):{\mathcal M}_{ij}\ne 0\}$. An edge $(v^+_i,v^-_j)\in {\mathbb E}$ is called a $*$-edge, if ${\mathcal M}_{ij}=*$, and a $?$-edge, if ${\mathcal M}_{ij}=?$. For any ${\mathcal I}\subseteq {\mathcal J}_m$ (resp. ${\mathcal I}\subseteq {\mathcal J}_n$), ${\mathbb V}^-({\mathcal I})$ (resp. ${\mathbb V}^+({\mathcal I})$) denotes $\{v^-_i\in {\mathbb V}^-:i\in {\mathcal I}\}$ (resp. $\{v^+_i\in {\mathbb V}^+:i\in {\mathcal I}\}$). Moreover, for any ${\mathbb V}_1\subseteq {\mathbb V}^+$, ${\mathbb V}_2\subseteq {\mathbb V}^-$, ${\mathbb E}({{\mathbb V}_1,{\mathbb V}_2})$ denotes $\{(v^+,v^-)\in {\mathbb E}: v^+\in {\mathbb V}_1, v^-\in {\mathbb V}_2\}$. An equivalent bipartite graph form of Theorem \ref{SolutionKone} is given as follows.
			
			\begin{proposition}\label{bipartite_form}
				Under Assumption \ref{throughout_assumption}, GLRM$({\mathcal M}, 1)$ is completable, if and only if there exists ${\mathcal I}\subseteq {\mathcal J}_m^{n-1}$, such that ${\rm mt}\left( ({\mathbb V}^+, {\mathbb V}^-({\mathcal I}), {\mathbb E}({{\mathbb V}^+,{\mathbb V}^-({\mathcal I})}),{\mathcal M}) \right)=n-1$, and for each $v^-_i\in {\mathbb V}^-({\mathcal J}_m\backslash {\mathcal I})$, one of the following conditions holds:
				
				1) for every $*$-edge $(v^+_j,v^-_i)\in {\mathbb E}({\mathbb V}^+,\{v^-_i\})$, \\ ${\rm mt}\left(({\mathbb V}^+\backslash \{v^+_j\}, {\mathbb V}^-({\mathcal I}), {\mathbb E}({{\mathbb V}^+\backslash \{v^+_j\}, {\mathbb V}^-({\mathcal I})}),{\mathcal M})\right)< n-1$;
				
				2) there exists a $?$-edge $(v^+_j,v^-_i)\in {\mathbb E}({\mathbb V}^+,\{v^-_i\})$ such that ${\rm mt}\left(({\mathbb V}^+\backslash \{v^+_j\}, {\mathbb V}^-({\mathcal I}), {\mathbb E}({{\mathbb V}^+\backslash \{v^+_j\}, {\mathbb V}^-({\mathcal I})}),{\mathcal M})\right)=n-1$.
			\end{proposition}
			
			\begin{proof}  Based on Theorem \ref{SolutionKone}, the result follows from the fact that the generic rank of a pattern matrix equals the cardinality of a maximum matching of its associated bipartite graph \cite{Murota_Book}.
			\end{proof}
			
			Proposition \ref{bipartite_form} has reduced the GLRM(${\mathcal M},1$) to checking the existence of certain maximum matchings in the associated bipartite graphs, which is intuitional. Its computational complexity is almost the same as Theorem \ref{SolutionKone}.
			
			\begin{example}Consider $\mathcal M$ in Example \ref{exp3}. The bipartite graph $\mathbb G(\mathcal M)$ is illustrated in Fig. \ref{exp4fig}. Choose ${\mathcal I}=\{1,2\}$ such that $({\mathbb V}^+, {\mathbb V}^-({\mathcal I}), {\mathbb E}({{\mathbb V}^+,{\mathbb V}^-({\mathcal I})}),{\mathcal M})$ has a maximum matching of cardinality $2$. From Fig. \ref{exp4fig}, it can be found that, for $i=3$, $(v_3^+,v_3^-)$ is a $?$-edge, while the associated bipartite graph has a maximum matching $\{(v_1^+,v_1^-),(v_2^+,v_2^-)\}$ with cardinality 2; for $i=4$, $(v_1^+,v_4^-)$ is a $?$-edge, while the associated bipartite graph has a maximum matching $\{(v_2^+,v_2^-),(v_3^+,v_1^-)\}$ with cardinality 2. From Proposition \ref{bipartite_form}, GLRM$({\mathcal M},1)$ is completable, which is consistent with Example \ref{exp3}.
			\end{example}
			
			\begin{figure}
				\centering
				\includegraphics[width=1.8 in]{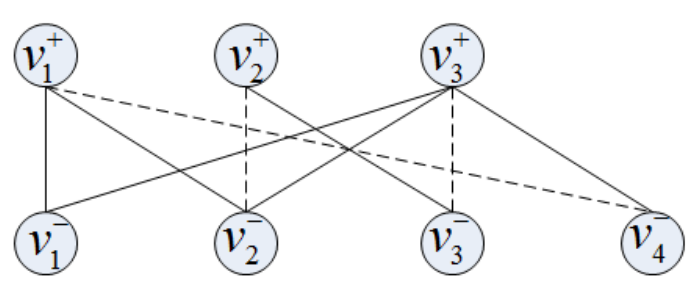}
				\caption{Bipartite graph representation of $\mathcal M$ in Example \ref{exp3}. Solid lines represent $*$-edges and dotted lines represent $?$-edges. }\label{exp4fig}
			\end{figure}

			\section{The case $k>1$} \label{sec-k}
			In this section, based on the results in the previous section, we study the $k$-GRR with $k>1$. We generalize the basis preservation principle and give some necessary/sufficient conditions for the completability of GLRM$({\mathcal M},k)$.
			\subsection{Sufficient condition}
			We first extend the basis preservation approach to a sufficient condition for the completability of GLRM$({\mathcal M},k)$.

			\begin{definition}[$k$-order preservable basis] Given a ${\mathcal M}\in \{0,*,?\}^{n\times m}$, a set ${\mathcal I}\in {\mathcal J}^{n-k}_m$ is called a $k$-order
				preservable basis of ${\mathcal M}$, if ${\rm grank} (\hat {\mathcal M}[:, {\mathcal I}])=n-k$.
			\end{definition}
			
			Roughly speaking, a $k$-order preservable basis indexes $n-k$ columns of $\hat {\mathcal M}$ (obtained from ${\mathcal M}$ by replacing all $?$ entries with $*$ ones) that has full column generic rank.	When $k=1$, the $k$-order preservable basis collapses to the preservable basis defined in Section \ref{sec-k-1}.
			
			\begin{theorem}\label{poly_verify}Under Assumption \ref{throughout_assumption}, suppose ${\mathcal I}\in {\mathcal J}^{n-k}_m$ such that ${\mathcal M}[:, {\mathcal I}]$ contains no $?$ entries and ${\rm grank }({\mathcal M}[:, {\mathcal I}])=n-k$. Then, GLRM$({\mathcal M},k)$ is completable, if and only if for each $i\in {\mathcal J}_{m}\backslash {\mathcal I}$
				\begin{equation} \label{formula-nonempty} |{\mathcal N}_{?i}|+{\rm grank} ({\mathcal M}[{\mathcal J}_n\backslash {\mathcal N}_{?i},{\mathcal I}])= |{\mathcal N}_i|+{\rm grank}( {\mathcal M}[{\mathcal J}_n\backslash {\mathcal N}_i,{\mathcal I}]).  \end{equation}
			\end{theorem}
			
			The proof of Theorem \ref{poly_verify} is postponed to Section \ref{proof}.

			\begin{proposition}(Sufficient condition for completability of GLRM$({\mathcal M},k)$) \label{suff-pro} Under Assumption \ref{throughout_assumption}, suppose there exists a $k$-order preservable basis of $\mathcal M$, denoted by $\mathcal I$. {If the condition in  Theorem \ref{poly_verify} (i.e., equality (\ref{formula-nonempty})) is satisfied with $\mathcal M$ replaced by ${\mathcal M}(\mathcal I)$, where ${\mathcal M}({\mathcal I})$ is defined in (\ref{basis-def}), then GLRM$({\mathcal M},k)$ is completable.}
			\end{proposition}
			
			\begin{proof}  On the basis of Theorem \ref{poly_verify}, the proof is similar to that for sufficiency of Theorem \ref{BasisPreservation}, and thus omitted. \end{proof}
			
The above proposition gives a sufficient condition for the completability of general $k$-GRR. It is easy to check that the condition in Theorem \ref{SolutionKone} is a special case of this proposition (the equivalence between the two conditions in Theorem \ref{SolutionKone} and the equality (\ref{formula-nonempty}) can be validated via Lemma \ref{rank-property} in Section \ref{proof}). In Theorem \ref{poly_verify}, {{for each fixed $\mathcal I$}}, condition (\ref{formula-nonempty}) can be verified simply via computing some maximum bipartite matchings (or generic ranks).  Concerning the complexity, for each ${\mathcal I}\in {\mathcal J}_{m}^{n-k}$, determine whether $\mathcal I$ is a $k$-order preservable basis incurs $O(n^{1.5}(n-k))$. Afterwards, for each $i\in {\mathcal J}_m\backslash {\mathcal I}$, checking formula (\ref{formula-nonempty}) again costs $O(n^{1.5}(n-k))$. In the worst case, there are $\binom{m}{n-k}$ $k$-order preservable bases that need to be verified. To sum up, the total computational complexity of Proposition \ref{suff-pro} is {$O(m^{\min\{n-k,m-n+k\}}n^{1.5}(n-k)(1+m))$, i.e., $O(m^{\min\{n-k,m-n+k\}}mn^{2.5})$}. 

			\begin{example} Consider the $k$-GRR with
			{\footnotesize$${\mathcal M}=\left[
				\begin{array}{cccc}
					* & ? & ? & * \\
					* & * & * & ? \\
					0 & 0 & 0 & * \\
				\end{array}
				\right]$$}and $k=2$.
				Let us select the first column of $\mathcal M$ as a $2$-order preservable basis, i.e., ${\mathcal I}=\{1\}$. Accordingly, the associated sets are ${\mathcal N}_{?2}=\{1\}$, ${\mathcal N}_{*2}=\{2\}$, ${\mathcal N}_{?3}=\{1\}$, ${\mathcal N}_{*3}=\{2\}$, ${\mathcal N}_{?4}=\{2\}$, and ${\mathcal N}_{*4}=\{1,3\}$. It can be verified that the condition in (\ref{formula-nonempty}) is satisfied for $i=2$ and $3$, while for $i=4$, the left-hand side of (\ref{formula-nonempty}) equals $2$ while the right-hand side equals $3$, leading to its violation. Therefore, GLRM$({\mathcal M},2)$ is incompletable. Indeed, one can easily find that the submatrix $M[\{2,3\},\{3,4\}]$ is generically invertible, verifying the assertion.
			\end{example}

			\subsection{Necessary condition}
			\begin{proposition}(Necessary condition for completability of GLRM$({\mathcal M},k)$) \label{necessary-cond}
				Under Assumption \ref{throughout_assumption}, if GLRM$({\mathcal M},k)$ is completable, then for every ${\mathcal I}\in {\mathcal J}^{n-k+1}_{n}$, GLRM$({\mathcal M}[{\mathcal I},:],1)$ is completable (which can be verified by Theorem \ref{SolutionKone}).
			\end{proposition}
			
			\begin{proof}  The required statement follows from the fact that every $n-k+1$ rows of $\mathcal M$ should permit a completion with rank no more than $n-k$.
			\end{proof}
			
			The above proposition gives a necessary condition for the completability of GLRM$({\mathcal M},k)$ by invoking Theorem \ref{SolutionKone}. Since verifying this condition has exponential complexity, from a practical view, one can select randomly a limited number of $n-k+1$ rows from $n$ rows of $\mathcal M$, and then verify the associated GLRM$({\mathcal M},1)$ using the randomized Algorithm \ref{alg1-theo4}. 

				\begin{remark} \label{suff-conje} We have implemented numerous simulations on different patterns $\mathcal M$ and found that the necessary condition in Proposition \ref{necessary-cond} is also {sufficient for those simulations}. We therefore conjecture that the condition in Proposition \ref{necessary-cond} is sufficient.  Proving or disproving this conjecture is left for future work.
				\end{remark}

				{\section{Applications to structured systems} \label{application}
This section {presents applications} of the proposed $k$-GRR framework to the resilience analysis of structured (descriptor) systems. Below, ${\mathcal P}({\mathcal M}_1,...,{\mathcal M}_n)$ denotes the set of tuples $(M_1,...,M_n)$ in which each $M_i\in {\mathcal P}({\mathcal M}_i)$.
					
					
					Consider a linear descriptor system described as \begin{equation}\label{plant} \begin{array}{c}
							E \dot x(t)=Ax(t)+Bu(t),\\
							y(t)=Cx(t)+Du(t),
					\end{array} \end{equation}
			
		where $t$ is the temporal indicator, $x(t)\in {\mathbb C}^{n}$, $u(t)\in {\mathbb C}^{q}$, and $y(t)\in {\mathbb C}^{p\times n}$ are respectively the state,  input, and output vectors. Accordingly, $E, A\in {\mathbb C}^{n\times n}$, $B\in {\mathbb C}^{n\times q}$, $C\in {\mathbb C}^{p\times n}$, and $D\in {\mathbb C}^{p\times q}$. System (\ref{plant}) is called {\emph{regular}} {whenever} $\lambda E-A$ is non-singular for some $\lambda\in {\mathbb C}$. {A vector $x\in {\mathbb C}^n$ is {\emph{admissible}} if there exists $u\in {\mathbb C}^q$ such that $Ax+Bu \in {\rm Im}(E)$.} Two vectors $x\in {\mathbb C}^n$ and $u\in {\mathbb C}^q$ are called {\emph{consistent}} if $Ax+Bu \in {\rm Im}(E)$. System (\ref{plant}) is {\emph{controllable}} (also known as R-controllability \cite{yip1981solvability}), if for any pair of admissible vectors $x_0, x_1\in {\mathbb C}^n$, there exist a finite time $T$ and a control input $u(t): [0,T]\to {\mathbb C}^q$ such that $x(0)=x_0$ and $x(T)=x_1$. System (\ref{plant}) is said to be {\emph{input-state observable}} (ISO), if for all consistent initial input $u(0)$ and state $x(0)$, $y(t)=0$ for $t\in [0,\infty)$ implies that $x(t)=0$ and $u(t)=0$ for $t\in [0,\infty)$.

					
					\begin{lemma}\cite{yip1981solvability,geerts1993invariant}
						The regular system (\ref{plant}) (or the triple $(E,A,B)$) is controllable if and only if
						${\rm rank} ([A-\lambda E, B])=n, \forall \lambda \in {\mathbb C}.$ A complex value $\lambda$ violating this condition is called an {\emph{uncontrollable mode}}. In addition, the regular system (\ref{plant}) (or the tuple $(E,A,B,C,D)$) is ISO,
						if and only if ${\rm rank}[A-\lambda E, B;C, D]=n+q$, $\forall \lambda \in {\mathbb C}.$ A complex value $\lambda$ violating this condition is called an invariant zero.
					\end{lemma}

When only the zero-nonzero patterns of $(E,A,B,C,D)$, denoted by $({\mathcal E},{\mathcal A},{\mathcal B}, {\mathcal C}, {\mathcal D})$, are available, we call the corresponding systems {\emph{structured (descriptor) systems}}. The pair $({\mathcal E},{\mathcal A})$ is called {\emph{structurally regular}} if there exist $\lambda\in {\mathbb C}$, $E\in {\mathcal P}({\mathcal E})$ and $A\in {\mathcal P}({\mathcal A})$ such that $\det (\lambda E-A)\ne 0$. Below, inspired by the rank $r$ resilience concept, we introduce the notions of robust controllability and robust ISO against structured perturbations.

					\begin{definition}
					Given $(E, A,B)$ and a triple of pattern matrices $(\delta {\mathcal E}, \delta {\mathcal A}, \delta {\mathcal B})$ specifying the zero-nonzero patterns of additive perturbations, $(E,A,B)$ is called robustly controllable w.r.t. $(\delta {\mathcal E}, \delta {\mathcal A},\delta {\mathcal B})$ if $(E+\delta E, A+\delta A, B+\delta B)$ is controllable for all $(\delta {\mathcal E}, \delta A, \delta B)\in {\mathcal P}(\delta {\mathcal E}, \delta {\mathcal A}, \delta {\mathcal B})$.\footnote{The perturbed system $(E+\delta E, A+\delta A, B+\delta B)$ fails to be controllable if either $(E+\delta E, A+\delta A)$ is not regular, or it has an uncontrollable mode.}
					\end{definition}	
					Similarly, we can define robust ISO of $(E,A,B,C,D)$  w.r.t. a perturbation pattern $(\delta {\mathcal E}, \delta {\mathcal A}, \delta {\mathcal B}, \delta {\mathcal C}, \delta {\mathcal D})$. The following result reveals that robust controllability and robust ISO of structured descriptor systems are generic properties. 
					
					\begin{proposition} \label{generic_control}
						Given two tuples $({\mathcal E}, {\mathcal A},{\mathcal B},{\mathcal C}, {\mathcal D})$ and $(\delta {\mathcal E}, \delta {\mathcal A}, \delta {\mathcal B}, \delta {\mathcal C}, \delta {\mathcal D})$, suppose that $({\mathcal E}, {\mathcal A})$ is structurally regular. The following statements hold:
						\begin{itemize}
							\item[i)] Either almost all $(E,A,B)\in {\mathcal P}({\mathcal E}, {\mathcal A},{\mathcal B})$ are robustly controllable w.r.t. $(\delta {\mathcal E}, \delta {\mathcal A},\delta {\mathcal B})$, or almost all $(E,A,B)\in {\mathcal P}({\mathcal E}, {\mathcal A},{\mathcal B})$ are not. 
							
							\item[ii)] Either almost all $(E, A,B,C,D)\in {\mathcal P}({\mathcal E}, {\mathcal A},{\mathcal B}, {\mathcal C}, {\mathcal D})$ are robustly ISO w.r.t. $(\delta {\mathcal E},\delta {\mathcal A},\delta {\mathcal B},\delta {\mathcal C},\delta {\mathcal D})$, or almost all $(E, A,B,C,D)\in {\mathcal P}({\mathcal E}, {\mathcal A},{\mathcal B}, {\mathcal C}, {\mathcal D})$ are not.
						\end{itemize}	
					\end{proposition}
					
					\begin{proof}  By setting $B=0,D=0$ and taking the transpose of $(E,A,C)$, ISO collapses to controllability. Hence, it suffices to prove statement ii). Let $p_*\doteq (p_{*1},\cdots, p_{*n_*})$ (resp. $p_?\doteq (p_{?1},\cdots, p_{?n_?})$) be a vector consisting of parameters for the nonzero entries in $({\mathcal E}, {\mathcal A},{\mathcal B}, {\mathcal C}, {\mathcal D})$ (resp. $(\delta {\mathcal E}, \delta {\mathcal A},\delta {\mathcal B}, \delta {\mathcal C}, \delta {\mathcal D})$). The respective realizations are denoted by $(E(p_*), A(p_*),B(p_*),C(p_*),D(p_*))$ and $(\delta E(p_?),\delta A(p_?),\delta B(p_?*),\delta C(p_?),\delta D(p_?))$.
						Define two matrix pencils $X(p_*,\lambda)\doteq[A(p_*)-\lambda E(p_*), B(p_*);C(p_*), D(p_*)]$ and $Y(p_*,\lambda)\doteq A(p_*)-\lambda E(p_*)$.
						Let $y(\lambda,p_*,p_?)=\det Y(p_*,p_?,\lambda)=\sum_{i=0}^{r}\beta_i(p_*,p_?)\lambda^i$ be a polynomial of $\lambda$ for some $r\ge 0$, where $\beta_i(p_*,p_?)$ are polynomials of $p_*,p_?$ ($i=0,...,r$). Consider two systems of polynomial equations in the variables $p_?$ and $\lambda$ as
\begin{align} \beta_i(p_*,p_?)&=0, \ \forall \  i \in \{0,...,r\},\label{non_feasible_constraint_control_regularity}\\
\det X(p_*,p_?,\lambda)[{\mathcal I}_1, :]&=0, \ \forall \  {\mathcal I}_1 \in {\mathcal J}_{n+p}^{n+q}. \label{non_feasible_constraint_control}
\end{align}
It turns out that $(E(p_*), A(p_*),B(p_*),C(p_*),D(p_*))$ is not robustly ISO w.r.t. $(\delta {\mathcal E}, \delta {\mathcal A},\delta {\mathcal B}, \delta {\mathcal C}, \delta {\mathcal D})$ if and only if either (\ref{non_feasible_constraint_control_regularity}) has a solution for $p_?\in {\mathbb C}^{n_?}$, or (\ref{non_feasible_constraint_control}) in the variables $p_?$ and $\lambda$ has a solution for $(p_?,\lambda)\in {\mathbb C}^{n_?}\times {\mathbb C}$.
						Note that (\ref{non_feasible_constraint_control}) is similar to (\ref{non_feasible_constraint}) except for the additional variable $\lambda$. Therefore, following the similar argument to the proof of Theorem \ref{feasibility_theorem}, it turns out that whether (\ref{non_feasible_constraint_control}) has a solution for $p_?\in {\mathbb C}^{n_?}$ and $\lambda \in {\mathbb C}$ depends on a condition similar to (\ref{finalcond}), whose satisfaction is generic for $p_*\in {\mathbb C}^{n_*}$; similar for the solvability of (\ref{non_feasible_constraint_control_regularity}). If either (\ref{non_feasible_constraint_control_regularity}) or (\ref{non_feasible_constraint_control}) is solvable for almost all $p_*\in {\mathbb C}^{n_*}$, then the corresponding realizations are not robustly ISO w.r.t. $(\delta {\mathcal E},\delta {\mathcal A},\delta {\mathcal B},\delta {\mathcal C},\delta {\mathcal D})$. Otherwise, for almost all $p_*\in {\mathbb C}^{n_*}$, both (\ref{non_feasible_constraint_control_regularity}) and (\ref{non_feasible_constraint_control}) are unsolvable, leading to the robust ISO of the corresponding realizations w.r.t. $(\delta {\mathcal E},\delta {\mathcal A},\delta {\mathcal B},\delta {\mathcal C},\delta {\mathcal D})$.  
					\end{proof}
					
		
				\begin{remark}\label{matrix-pencil-completion}
			The proof of Proposition \ref{generic_control} can be generalized to show the genericity of the completability of the following {\emph{low-rank matrix pencil}} in the space for parameters of observed entries in $X$ indexed by $\Omega\subseteq {\mathcal J}_n\times {\mathcal J}_q$:
			
			
			\begin{equation} \label{lowrankproblem-pencil}
			\begin{array}{c}
			\begin{aligned}
			{\rm{find}} \quad & \lambda\in {\mathbb C},  X\in {\mathbb C}^{n\times q} \\
			{\rm {s.t.}} \quad & X_{ij}=x_{ij}, \ \forall (i,j)\in \Omega, \\
			\quad &   {\rm {rank}}  (\lambda Y+ X)\le n-k,
			\end{aligned}
			\end{array}
			\end{equation}
			where $Y\in {\mathbb C}^{n\times q}$ is a given constant, $\Omega$ is the set of indices for the observed entries in $X$, and $x_{ij}$ denotes the observed entry at the $(i,j)$th entry of $X$.  It can be seen that when $Y=0$, the above problem reduces to the conventional LRMC problem.
		\end{remark}
				
Proposition \ref{generic_control} naturally induces the following concepts. $({\mathcal E}, {\mathcal A},{\mathcal B})$ is said to be {\emph{structurally robustly controllable}} w.r.t. $(\delta {\mathcal E}, \delta {\mathcal A}, \delta {\mathcal B})$ if almost all $(E,A,B)\in {\mathcal P}({\mathcal E}, {\mathcal A},{\mathcal B})$ are robustly controllable w.r.t. $(\delta {\mathcal E}, \delta {\mathcal A},\delta {\mathcal B})$. Similarly, we can define {\emph{structural robust ISO}} w.r.t. a perturbation pattern. These concepts generalize the notion {\emph{perturbation-tolerant structural controllability}} in \cite{full-version-tac} to descriptor systems and extend it to the ISO property. {Note that \cite{full-version-tac} relies on the controllability matrix to justify this concept, which is not applicable to descriptor systems since a similar controllability matrix polynomially dependent on the system matrices is generically inaccessible for descriptor systems.} When $\delta {\mathcal E},\delta {\mathcal A},\delta {\mathcal B}$ are zero matrices (i.e., with no perturbations), Proposition \ref{generic_control} justifies the generic property of controllability of structured descriptor systems.


While the $k$-GRR framework does not provide complete criteria for the proposed structural robust controllability/ISO, it does establish necessary and sufficient conditions for the existence of the zero uncontrollable mode and the zero invariant zero against structured perturbations. For ease of description, we focus on normal systems, i.e., $E=I_n$.

					\begin{proposition} \label{control_system_application}
						Consider two tuples $({\mathcal A},{\mathcal B}, {\mathcal C}, {\mathcal D})$ and $(\delta {\mathcal A}, \delta {\mathcal B}, \delta {\mathcal C}, \delta {\mathcal D})$ associated with a normal system (\ref{plant}) with $E=I_n$. For almost all $(A,B)\in {\mathcal P}({\mathcal A},{\mathcal B})$, there exists $(\delta A,\delta B)\in {\mathcal P}(\delta {\mathcal A},\delta {\mathcal B})$ such that the perturbed system $(A+\delta A, B+\delta B)$ has a zero uncontrollable mode, if and only if $[{\mathcal A},{\mathcal B}]$ is structurally rank $n-1$ completable against $[\delta {\mathcal A},\delta {\mathcal B}]$.
Moreover, for almost all $(A,B,C,D)\in {\mathcal P}({\mathcal A},{\mathcal B}, {\mathcal C}, {\mathcal D})$, there exists $(\delta A,\delta B, \delta C, \delta D)\in {\mathcal P}(\delta {\mathcal A},\delta {\mathcal B}, \delta {\mathcal C}, \delta {\mathcal D})$ such that the perturbed system $(A+\delta A, B+\delta B, C+\delta C, D+\delta D)$ has an zero invariant zero, if and only if $[{\mathcal A},{\mathcal B};
									{\mathcal C},{\mathcal D}]$ is structurally rank $n+q-1$ completable against $[\delta{\mathcal A},\delta{\mathcal B};
									\delta{\mathcal C},\delta{\mathcal D}]$.
					\end{proposition}
					
					\begin{proof}
						The result is immediate from the definitions of zero uncontrollable mode and zero invariant zeros.
					\end{proof}


	The above result partially extends the criteria for the perturbation-tolerant structural controllability in \cite{full-version-tac} from single-input systems to the multi-input case. In what follows, we show that the $k$-GRR may help to characterize zero {\emph{structurally fixed mode}} (SFM).

				\begin{definition}\cite[sec 6]{Murota_Book} System (\ref{plant}) with $E=I_n$, denoted by $(A,B,C)$, has a fixed mode $\lambda_0\in {\mathbb C}$ w.r.t. ${\mathcal K}\in \{0,*\}^{m\times p}$, if ${\rm rank}(A-\lambda_0I+BKC)< n$ holds $\forall K\in {\mathcal P}(\mathcal K)$.  The corresponding normal structured system (\ref{plant}), denoted by $({\mathcal A},{\mathcal B}, {\mathcal C})$, has no structurally fixed mode w.r.t. ${\mathcal K}$, if almost all $(A,B,C)\in {\mathcal P}({\mathcal A},{\mathcal B}, {\mathcal C})$ have no fixed mode w.r.t. ${\mathcal K}$. 				
				\end{definition}
				
				
In particular,  $({\mathcal A},{\mathcal B}, {\mathcal C})$ has no zero SFM, if almost all $({\mathcal A},{\mathcal B}, {\mathcal C})$ has no zero fixed mode $\lambda_0=0$. The absence of fixed modes is necessary and sufficient for {\emph{arbitrary pole placement with output feedback using dynamic compensators}} \cite{wang1973stabilization}. 
				By the Schur complement, ${\rm rank}(A+BKC)< n$, if and only if (see \cite[Theo  6.5.1]{Murota_Book})
				$$M(A,B,C,K)\doteq \left[\begin{array}{ccc}
					0 & A & B \\
					I_p & C & 0 \\
					K & 0 & I_q
				\end{array}
				\right]$$
				fails to have full column rank. Therefore, $(A,B,C)$ has no zero fixed mode, if and only if $M(A,B,C,K)$ is rank $(p+n+m)$ resilient against ${\mathcal K}$. Let ${\mathcal M}({\mathcal A},{\mathcal B}, {\mathcal C}, {\mathcal K})$ be the pattern matrix corresponding to $M(A,B,C,K)$. Correspondingly, {\emph{ $({\mathcal A},{\mathcal B}, {\mathcal C})$ has no zero SFM w.r.t. ${\mathcal K}$, if and only if ${\mathcal M}({\mathcal A},{\mathcal B}, {\mathcal C}, {\mathcal K})$ is structurally rank $(p+n+q)$ resilient against ${\mathcal K}$. }} It follows that Theorem \ref{SolutionKone} can provide a necessary and sufficient condition for the absence of zero SFM.

				\section{Proofs of Theorems \ref{BasisPreservation},  \ref{BasisApproach} , and \ref{poly_verify}} \label{proof}
				This section gives proofs for Theorems \ref{BasisPreservation}, \ref{BasisApproach}, and \ref{poly_verify}. In our proofs, the following properties of the rank function will be used frequently.
				
				\begin{lemma}\cite[Sec 2.3]{Murota_Book} \label{rank-property} Let $M\in {\mathbb C}^{n\times m}$.  For ${\mathcal I}_1, {\mathcal I}_2\subseteq {\mathcal J}_m$ satisfying ${\rm rank} (M[:,{\mathcal I}_2]) =|{\mathcal I}_2|$, the following properties hold.
					
					1) If ${\mathcal I}_1\subseteq {\mathcal I}_2$ then ${\rm rank}(M[:,{\mathcal I}_1])=|{\mathcal I}_1|$.
					
					2) If ${\rm rank} (M[:,{\mathcal I}_1]) =|{\mathcal I}_1|$ and $|{\mathcal I}_1|<|{\mathcal I}_2|$,  there is ${\mathcal I}_3\subseteq {\mathcal I}_2\backslash {\mathcal I}_1$ and $|{\mathcal I}_1\cup{\mathcal I}_3|=|{\mathcal I}_2|$ making ${\rm rank}(M[:,{\mathcal I}_1\cup {\mathcal I}_3])=|{\mathcal I}_2|$.
					
				\end{lemma}
				
				\subsection{Proof of Theorem \ref{BasisPreservation}}
				{\bf Proof for Sufficiency of Theorem \ref{BasisPreservation}}: Let $p_*$ be the set of parameters for $*$ entries in $\mathcal M$, $\bar p_*$ for $?$ entries in ${\mathcal M}[:, {\mathcal I}]$, and $\bar p_{?}$ for $?$ entries in ${\mathcal M}[:, {\mathcal J}_m\backslash {\mathcal I}]$, whose numbers of elements are denoted by $n_1$, $n_2$, and $n_3$, respectively. If
				GLRM$({\mathcal M}({\mathcal I}),1)$ is completable, then for almost all $(p_*,\bar p_*)$ in ${\mathbb C}^{n_1+n_2}$, there exists $\bar p_{?}\in {\mathbb C}^{n_3}$ making the corresponding matrix completions row rank deficient. It is clear that all $p_*$ satisfying this condition forms a set ${\mathcal P}^*$ with nonzero measure in ${\mathbb C}^{n_1}$, as otherwise the set of values for $(p_*,\bar p_*)$ will have zero measure in ${\mathbb C}^{n_1+n_2}$. Thereby, for any $p_*\in {\mathcal P}^*$, there exist corresponding $\bar p_* \in {\mathbb C}^{n_2}$ and $\bar p_{?} \in {\mathbb C}^{n_3}$ making the associated completions row rank deficient, proving the completability of GLRM$({\mathcal M}, 1)$. $\hfill\square$

				Our proof of necessity can be partitioned into three parts.
				
				\begin{lemma} \label{mid1} Under Assumption \ref{throughout_assumption}, if GLRM$({\mathcal M},1)$ is completable, then GLRM$({\mathcal M},1)$ subject to the additional constraint that the matrix completion has rank $n-1$ is completable.
				\end{lemma}
				
				\begin{proof}  Under Assumption \ref{throughout_assumption}, for almost all $M\in {\mathcal P}({\mathcal M}^{o})$, one has ${\rm rank} (M)= n$. Since GLRM(${\mathcal M},1$) is completable, for almost all $M\in {\mathcal P}({\mathcal M}^{o})$ satisfying ${\rm rank}(M)=n$, there is a completion $X$ of $M$ with  ${\rm rank}(X) \le n-1$. Suppose ${\rm rank}(X) < n-1$. Then, changing $X$ into $M$ by modifying one entry of $X$ into zero (corresponding to setting some $?$ entry to be zero) at a time changes the rank by at most one at each step. Therefore, a completion of rank $n-1$ can be obtained, for almost all $M\in {\mathcal P}({\mathcal M}^{o})$.
				\end{proof}

				\begin{lemma} \label{mid2} Under Assumption \ref{throughout_assumption}, if GLRM$({\mathcal M},1)$ is completable, then there exists ${\mathcal I}\in {\mathcal J}^{n-1}_m$, such that GLRM $({\mathcal M}({\mathcal I}),1)$ is completable.
				\end{lemma}

	\begin{proof}  From Lemma \ref{mid1}, under the proposed condition, there generically exists a completion of ${\mathcal M}$ with rank $n-1$. Denote such a completion by $X(p_*,p_?)$, namely, ${\rm rank}(X(p_*,p_?))=n-1$.
	Then, there exists an ${\mathcal K}\in {\mathcal J}^{n-1}_m$ satisfying
${\rm rank}(X(p_*,p_?)[:,{\mathcal K}])=n-1.$
	Under this condition, it is not hard to see that the constraint ${\rm rank} (X(p_*,p_?))= n-1$ is equivalent to
	\begin{equation} \label{constraint}
		{\det (X(p_*,p_?)[:,{\mathcal K}\cup \{i\}])}=0, \ \forall i \in {\mathcal J}_m\backslash {\mathcal K}.
	\end{equation}

{Note (\ref{constraint}) induces a system of polynomial equations in the variables $p_?$, which should have a solution for almost all $p_*\in {\mathbb C}^{n_*}$ by the completability of the corresponding GLRM. Now regard (\ref{constraint}) as an undetermined system of $|{\mathcal J}_m\backslash {\mathcal K}|=m-n+1$ polynomial equations in $n_?+n_*$ variables $p_*\cup p_?$. From Lemma \ref{system_polynomial}, the set of all complex solutions is an algebraic variety with dimension at least $n_?+n_*-(m-n+1)$. It then follows that at most $m-n+1$ variables in $p_?\cup p_*$ are algebraic over the field obtained by adjoining the remaining $n_?+n_*-(m-n+1)$ variables to $\mathbb C$ (that is, at most $m-n+1$ variables in $p_?\cup p_*$ can be written as solutions of the system of polynomial equations with coefficients being the remaining $n_?+n_*-(m-n+1)$ algebraically independent variables and numbers in $\mathbb C$; see footnote \ref{algebraic_dimension_foot1}). As $m-n+1$ variables locate at most $m-n+1$ columns of ${\mathcal M}$, it turns out that there are at least $n-1$ ($=m-(m-n+1)$) columns in ${\mathcal M}$, such that regarding the $?$ entries therein as generic entries ($*$ entries) will not sacrifice the completability of the corresponding GLRM. That is, there is some ${\mathcal I}\subseteq {\mathcal J}_{m}^{n-1}$ such that GLRM $({\mathcal M}({\mathcal I}),1)$ is completable.
			}\end{proof}
				
				
				{\bf Proof of Necessity of Theorem \ref{BasisPreservation}}:  It remains to prove that the set $\mathcal I$ defined in Lemma \ref{mid2} could be {\emph{restricted to a preservable basis}}, i.e., ${\rm grank} (\hat {\mathcal M}[:,{\mathcal I}])=n-1$, recalling that $\hat {\mathcal M}$ is obtained from $\mathcal M$ by replacing all $?$ entries with $*$ entries.\footnote{Notice that the set ${\mathcal K}$ in the proof of Lemma \ref{mid2} does not necessarily coincide with ${\mathcal I}$.}
				
				Suppose that GLRM$({\mathcal M}({\mathcal I}), 1)$ is completable, with ${\mathcal I}\in {\mathcal J}^{n-1}_m$. If ${\rm grank}(\hat {\mathcal M}[:,{\mathcal I}])=n-1$, the proof is finished. Now consider the case ${\rm grank} (\hat {\mathcal M}[:,{\mathcal I}])=n_0< n-1$. From Lemma \ref{rank-property}, there is ${\mathcal I}_0\subseteq {\mathcal I}$ with $|{\mathcal I}_0|=n_0$ such that
				${\rm grank}(\hat {\mathcal M}[:,{\mathcal I}_0])=n_0$. As GLRM$({\mathcal M}({\mathcal I}), 1)$ is completable, so is GLRM$({\mathcal M}({\mathcal I}_0),1)$. On the other hand, let $X$ be a completion of ${\mathcal M}({\mathcal I})$. Then, it generally holds ${\rm Im}(X[:,{\mathcal I}_0])= {\rm Im}(X[:,{\mathcal I}])$, followed by
				\begin{equation}\label{rank_eq} {\rm rank}(\left[X[:,{\mathcal I}],X[:,{\mathcal J}_m\backslash {\mathcal I}]\right])={\rm rank}(\left[X[:,{\mathcal I}_0],X[:,{\mathcal J}_m\backslash {\mathcal I}]\right]).\end{equation}
				This indicates that upon defining ${\mathcal M}_1\doteq [\hat {\mathcal M}[:,{\mathcal I}_0],{\mathcal M}[:,{\mathcal J}_m\backslash {\mathcal I}]]$ (i.e., deleting the columns indexed by ${\mathcal I}\backslash {\mathcal I}_0$ from $\mathcal M$), GLRM$({\mathcal M}_1,1)$ is also completable.  Using Lemma \ref{mid2} on $({\mathcal M}_1,1)$, there is a set ${\mathcal I}_1$ with $|{\mathcal I}_1|=n-1$, such that $({\mathcal M}_1({\mathcal I}_1),1)$ is completable. Define ${\mathcal M}_{2}\doteq[\hat {\mathcal M}_1[:,{\mathcal J}_{n_0}],{\mathcal M}_1[:,{\mathcal J}_{m_1}\backslash {\mathcal I}_1]]$, with $m_1$ the number of columns of ${\mathcal M}_1$, recalling that $\hat {\mathcal M}_1[:, {\mathcal J}_{n_0}]=\hat {\mathcal M}[:,{\mathcal I}_0]$. If ${\rm grank} (\hat {\mathcal M}_1[:, {\mathcal I}_1\cup {\mathcal J}_{n_0}])={\rm grank} (\hat {\mathcal M}_1[:, {\mathcal J}_{n_0}]) = n_0$, then using the same reasoning as (\ref{rank_eq}), we get GLRM$({\mathcal M}_{2},1)$ is also completable. Meanwhile, upon letting $\bar {\mathcal I}_1\subseteq {\mathcal J}_m$ such that $\hat {\mathcal M}[:,\bar {\mathcal I}_1]=\hat {\mathcal M}_1[:,{\mathcal I}_1\cup {\mathcal J}_{n_0}]$,
				GLRM$({\mathcal M}(\bar {\mathcal I}_1),1)$ is completable, too.  	 We can apply the above procedure from ``$0\to 1$'' to ``$k\to k+1$'', for $k=0,1,2,\cdots$, until ${\mathcal I}_{\bar k}$ emerges for some ${\bar k}\in {\mathbb N}^+$, such that ${\rm grank}(\hat {\mathcal M}_{\bar k}[:, {\mathcal I}_{\bar k}\cup {\mathcal J}_{n_0}])> n_0$ meanwhile GLRM$({\mathcal M}_{\bar k},1)$ is completable. 		 Note that such a ${\mathcal I}_{\bar k}$ must exist, as otherwise ${\rm grank}(\hat {\mathcal M})=n_0<n={\rm grank}({\mathcal M}^{p})$, causing a contradiction to Assumption \ref{throughout_assumption}. Let $\bar {\mathcal I}_{\bar k}$ be such that $\hat {\mathcal M}[:,\bar {\mathcal I}_{\bar k}]=\hat {\mathcal M}_{\bar k}[:, {\mathcal I}_{\bar k}\cup {\mathcal J}_{n_0}]$. 	 Then, it can be seen GLRM$({\mathcal M}(\bar {\mathcal I}_{\bar k}),1)$ is completable because of the completability of  GLRM$({\mathcal M}_{\bar k},1)$. Consequently, there is a subset ${\mathcal I}^1_0\subseteq {\bar {\mathcal I}_{\bar k}}$, such that ${\rm grank}(\hat {\mathcal M}[:,\bar {\mathcal I}_{\bar k}])={\rm grank}(\hat {\mathcal M}[:,{\mathcal I}^1_0])=|{\mathcal I}^1_0|>n_0$. The completability of GLRM$({\mathcal M}({\mathcal I}^1_0),1)$ follows directly from that of GLRM$({\mathcal M}(\bar {\mathcal I}_{\bar k}),1)$.

				To sum up, for any ${\mathcal I}_0\subseteq {\mathcal J}_m$ such that GLRM$({\mathcal M}({\mathcal I}_0),1)$ is completable while ${\rm grank} (\hat {\mathcal M}[:,{\mathcal I}_0])=|{\mathcal I}_0|<n$, by means of the above procedure, we can always find a ${\mathcal I}^1_0\subseteq {\mathcal J}_m$, such that GLRM$({\mathcal M}({\mathcal I}_0^1),1)$ is completable, while
				${\rm grank} (\hat {\mathcal M}[:,{\mathcal I}^1_0])=|{\mathcal I}^1_0|\ge |{\mathcal I}_0|+1$. Hence, after repeating this reasoning at most $n-1-n_0$ times, we can always find a ${\mathcal I}_0^t$ ($t\le n-1-n_0$) with ${\rm grank} (\hat {\mathcal M}[:,{\mathcal I}^t_0])=n-1$ (i.e., ${\mathcal I}_0^t$ is a preservable basis), such that GLRM$({\mathcal M}({\mathcal I}_0^t),1)$ is completable.
				$\hfill\square$

				\subsection{Proof of Proposition \ref{BasisApproach}}
				To prove Proposition \ref{BasisApproach}, the following lemma is needed. 
				
				\begin{lemma}\label{leftnullspace} Given an $p\times q$ matrix $H$, let $T$ consist of a set of linearly independent {\emph{row}} vectors spanning the left null space of $H$. Then, for any ${\mathcal I}\subseteq {\mathcal J}_p$, $T[:,{\mathcal I}]$  is of full row rank, if and only if $H[{\mathcal J}_p \backslash {\mathcal I},:]$ is of full row rank.
				\end{lemma}
				
				\begin{proof} 
Without losing any generality, consider the case ${\mathcal I}=\{1,...,l\}$, $l\le p$. Partition $H$ as
					$$H=\left[
					\begin{array}{c}
						H_1 \\
						H_2 \\
					\end{array}
					\right]\doteq \left[
					\begin{array}{c}
						H[{\mathcal I},:]\\
						H[{\mathcal J}_p\backslash {\mathcal I},:] \\
					\end{array}
					\right],
					$$
					and $T$ as $T=[T_1,T_2]\doteq \left[T[:,{\mathcal I}],T[:,{\mathcal J}_p\backslash {\mathcal I}]\right]$.
					
					$\Rightarrow:$ Suppose that $H_2$ is of row rank deficient. Then, there exists a nonzero matrix $\hat T_2$ of dimension ${r\times (p-l)}$ for some $r\ge 1$, satisfying $\hat T_2 H_2=0$, which leads to $[0_{r\times l}, \hat T_2]H=0$. As the rows of $T$ form a basis for the left null space of $H$, there must exist a nonzero matrix $K$ such that
					$[0_{r\times l}, \hat T_2]=KT=K[T_1,T_2],$ which leads to $KT_1=0$, indicating that $T_1$ is of row rank deficient.
					
					$\Leftarrow:$ Suppose that $T_1$ is of row rank deficient. Then, there exists a nonzero matrix $K$ making $KT_1=0_{r\times l}$, for some $r\ge 1$. As a result,
					$KTH=K[T_1,T_2]H=[0,KT_2H_2]=0.$ Note that $KT_2\ne 0$, as otherwise $KT=K[T_1,T_2]=0$, contradictory to the full row rank of $T$. Therefore,  $H_2$ is of row rank deficient.
				\end{proof}
				
				{\bf Proof of Proposition \ref{BasisApproach}:}  Let $X$ be a completion of $\mathcal M$. As ${\rm grank}({\mathcal M}[:, {\mathcal I}])=n-1$ and there is no $?$ entry in $X[:, {\mathcal I}]$, it generally holds that ${\rm rank} (X[:, {\mathcal I}])=n-1$. Let $q$ be an $n$-vector spanning the left null space of $X$. From Lemma \ref{leftnullspace}, $q_j\ne 0$ if and only if $j\in {\mathcal B}^*$ for almost all $X[:,{\mathcal I}]\in {\mathcal P}(\mathcal M[:,{\mathcal I}])$. Hence, it generically holds $q^{\intercal} X[:,\{i\}]=\sum \nolimits_{j\in {\mathcal B}^*} q_jX_{ji}$, for each $i\in {\mathcal J}_m\backslash {\mathcal I}$.
				
				{\bf Sufficiency}: If condition 1) is satisfied, by setting ${X}_{ji}=0$ whenever $j\in {\mathcal N}_{?i}$, one can always make $q^{\intercal} {X}[:,\{i\}]=0$ for each $i\in {\mathcal J}_m\backslash {\mathcal I}$. Thereby $q^{\intercal }{X}=0$, making ${\rm rank}({X})\le n-1$.	If condition 2) is satisfied, let ${\mathcal L}_i\doteq {\mathcal B}^*\cap {\mathcal N}_{?i}$. By setting $X_{ki}=-1/q_k\sum \nolimits_{j\in {{\mathcal B}^*\cap {\mathcal N}_{*i}}}q_jX_{ji}$ for some $k\in {\mathcal L}_i$, and $X_{ki}=0$ for all $k\in {\mathcal N}_{?i}\backslash \{k\}$, we have
				$q^{\intercal}X[:,\{i\}]=\sum \nolimits_{i\in {\mathcal B}^*} q_iX_{ji}=0,$ for each $i\in {\mathcal J}_m\backslash {\mathcal I}$,
				which means $q^{\intercal}X=0$, leading to ${\rm rank}(X) \le n-1$.
				
				{\bf Necessity:} If neither condition 1) nor condition 2) is satisfied, then there exists some $i\in {\mathcal J}_m\backslash {\mathcal I}$ such that ${\mathcal B}^*\cap {\mathcal N}_{*i}\neq \emptyset$, but ${\mathcal B}^*\cap {\mathcal N}_{?i}= \emptyset$. This leads to
				$q^{\intercal}X[:,\{i\}]=\sum \nolimits_{j\in {\mathcal B}^*\cap {\mathcal N}_{*i}} q_jX_{ji},$
				which is generally not zero, as $\{q_j:j\in {\mathcal B}^*\cap {\mathcal N}_{*i}\}$ are unique (up to scaling) whiles $\{X_{ji}:j\in {\mathcal B}^*\cap {\mathcal N}_{*i}\}$ are independent. Consequently, $X$ is generically of full row rank.
				
				\subsection{Proof of Theorem \ref{poly_verify}}

				Some intermediate results are given for the proof of Theorem \ref{poly_verify}. The following lemma gives a null-space based condition for an $n\times m$  matrix to have rank $n-k$.
				
				\begin{lemma} \label{leftspaceprinciple}
					Let $T=[T_1,T_2]\in {\mathbb C}^{n\times m}$, with $T_1$ and $T_2$ having dimensions of $n\times (n-k)$ and $n\times (m-n+k)$, respectively, $k\in {\mathbb N}$. In addition, $T_1$ is
					of full column rank. Suppose $\Gamma$ consists of a set of linearly independent row vectors spanning the left null space of $T_1$. Then, ${\rm rank} (T)= n-k$, if and only
					if $\Gamma T_2=0$.
				\end{lemma}
				
				{\begin{proof}
						Sufficiency: As $\Gamma T_2=0$, ${\rm Im} (T_2)\subseteq {\rm null}(\Gamma)$. Therefore, ${\rm Im} (T_2) \subseteq {\rm Im}(T_1)$ as ${\rm null}(\Gamma)={\rm Im}(T_1)$, which indicates ${\rm rank} (T)= {\rm rank} (T_1)=n-k$. Necessity: Assume that $\Gamma T_2\ne 0$. Then, there is a vector $x\in {\rm Im}(T_2)$ but $x\notin {\rm null}(\Gamma)$. The latter indicates $x\notin {\rm Im}(T_1)$. Hence, ${\rm rank}(T)> {\rm rank} (T_1)=n-k$, causing a contradiction.
				\end{proof}}
				
				\begin{lemma} \label{theorem_k_less} Under Assumption \ref{throughout_assumption}, suppose that there exists a set of columns of $\mathcal M$ indexed by
					${\mathcal I}\in {\mathcal J}^{n-k}_m$ such that ${\mathcal M}[:, {\mathcal I}]$ contains no $?$ entries and ${\rm grank }({\mathcal M}[:, {\mathcal I}])=n-k$.
					Moreover, define ${\mathcal B}=\{{\mathcal K}\subseteq {\mathcal J}^{n-k}_n: {\rm grank}( {\mathcal M}[{\mathcal K}, {\mathcal I}])=n-k\}$, and let ${\mathcal B}^*= \{{\mathcal J}_n\backslash \Omega: \Omega \in {\mathcal B}\}$. For $i\in {\mathcal J}_{m}\backslash {\mathcal I}$, define ${\rho}_i\doteq \max \limits_{w\in {{\mathcal B}^*}} |w\cap {\mathcal N}_i|$ and let
					${\mathcal S}_i=\{{\mathcal S}\subseteq {\mathcal N}_{?i}: \exists \ w\in {\mathcal B}^*, {\rm s.t.}\ {\mathcal S}\subseteq w, |{\mathcal S}|=\rho_i\},$with ${\mathcal N}_{?i}, {\mathcal N}_{*i}$, and ${\mathcal N}_{i}$ defined in Section \ref{approach}. GLRM$({\mathcal M},k)$ is completable, if and only if for each $i\in {\mathcal J}_{m}\backslash {\mathcal I}$, either $\rho_i=0$ or ${\mathcal S}_i\ne \emptyset$. 
				\end{lemma}

				{\begin{proof}
						Let $X(p_*,p_?)$ be a completion of $\mathcal M$ with parameters $p_*,p_?$. In the following, $p_*$ and $p_?$ will be dropped for notation simplicity. Let ${\bar \Gamma}\in {\mathbb C}^{k\times n}$ denote the matrix consisting of a set of linearly
						independent {\emph{row}} vectors spanning the left null space of $X[:,{\mathcal I}]$. From Lemmas \ref{rank-property} and \ref{leftnullspace}, we know  for each $w\in {\mathcal B}^*$, it generically holds
						\begin{equation}\label{full-rank} {\rm rank}(\bar \Gamma[:,w])=k.\end{equation}
						Using 1) of Lemma \ref{rank-property} on (\ref{full-rank}), we get ${\rm rank} (\bar \Gamma[:,{\mathcal N}_i])= \max \nolimits_{w\in {{\mathcal B}^*}} |w\cap {\mathcal N}_i|=\rho_i$.
						
						{\bf Sufficiency}:  We shall use Lema \ref{leftspaceprinciple}, i.e., by showing that $\bar \Gamma X[:,{\mathcal J}_m\backslash {\mathcal I}]=0$, to prove
						${\rm rank} (X)= {\rm rank}(X[:,{\mathcal I}])=n-k$.  If $\rho_i=0$, then ${\rm rank} (\bar \Gamma[:,{\mathcal N}_i])=\rho_i=0$, leading to $\bar \Gamma[:,{\mathcal N}_i]=0$. It follows that $\bar \Gamma X[:,\{i\}]=\bar \Gamma[:,{\mathcal N}_i] X[{\mathcal N}_i,\{i\}]=0$. Now suppose $\rho_i>0$ and there is an $S\in {\mathcal S}_i$ such that ${\mathcal S}_i\ne \emptyset$. From (\ref{full-rank}), we have ${\rm rank}({\bar \Gamma}[:,S])=\rho_i$. Note that for each $i\in {\mathcal J}_m\backslash {\mathcal I}$, $\bar \Gamma X[:,\{i\}]= \bar \Gamma[:, {\mathcal N}_i]X[{\mathcal N}_i,\{i\}]=\bar \Gamma[:,S]X[S,\{i\}]+\bar \Gamma[:,{\mathcal N}_i\backslash S]X[{\mathcal N}_i\backslash S,\{i\}]$.
						 As ${\rm rank}({\bar \Gamma}[:,{\mathcal N}_i])={\rm rank}({\bar \Gamma}[:,S])=\rho_i$,  we have ${\rm Im}(\bar \Gamma[:,{\mathcal N}_i\backslash S]) \subseteq {\rm Im}(\bar \Gamma[:,S])$. Since $S\subseteq {\mathcal N}_{?i}$, setting $X[S,\{i\}] =-((\bar \Gamma[:,S])^{\intercal}\bar \Gamma[:,S])^{-1}(\bar \Gamma[:,S])^{\intercal} \bar \Gamma[:,{\mathcal N}_i\backslash S]X[{\mathcal N}_i\backslash S,\{i\}]$ yields $\bar \Gamma X[:,\{i\}]=\bar \Gamma[:,S]X[S,\{i\}]+\bar \Gamma[:,{\mathcal N}_i\backslash S]X[{\mathcal N}_i\backslash S,\{i\}]=$0, $\forall X[{\mathcal N}_{*i},\{i\}]\in {{\mathbb C}^{|{\mathcal N}_{*i}|}}$, where the $?$ entries in $X[{\mathcal N}_i\backslash S,\{i\}]$ can be arbitrarily assigned.
						Hence, by Lemma \ref{leftspaceprinciple}, ${\rm rank} (X)= n-k$, which proves the sufficiency.
						
						{\bf Necessity}: Again from Lemma \ref{leftspaceprinciple}, it is necessary that $\bar \Gamma X[:,\{i\}]=0$ holds for each
						$i\in {\mathcal J}_m\backslash {\mathcal I}$ for ${\rm rank } (X)= n-k$ to be true.
						Notice that
						\begin{equation} \label{equaltozero} \begin{array}{c}
								\bar \Gamma X[:,\{i\}]= \bar \Gamma[:, {\mathcal N}_i]X[{\mathcal N}_i,\{i\}]= \\ \bar \Gamma[:,{\mathcal N}_{?i}]X[{\mathcal N}_{?i},\{i\}]+\bar \Gamma[:,{\mathcal N}_{*i}]X[{\mathcal N}_{*i},\{i\}].
								
							\end{array}
						\end{equation}Suppose by the sake of contradiction that there is an integer $i\in {\mathcal J}_m\backslash {\mathcal I}$ such that ${\mathcal S}_i=\emptyset$ and $\rho_i>0$.  Then, there is at least one column of $\bar \Gamma[:,{\mathcal N}_{*i}]$, denoted by $x$, such that
						${\rm Im} (x)\notin  {\rm Im} (\bar \Gamma[:,{\mathcal N}_{?i}])$. Indeed, if the range space of every column of $\Gamma[:,{\mathcal N}_{*i}]$ is contained in ${\rm Im}(\bar \Gamma[:,{\mathcal N}_{?i}])$, then ${\rm rank}(\bar \Gamma[:,{\mathcal N}_{i}])={\rm rank} (\bar \Gamma[:,{\mathcal N}_{?i}\cup {\mathcal N}_{*i}])={\rm rank} (\bar \Gamma[:,{\mathcal N}_{?i}])={\rho}_i$, which indicates ${\mathcal S}_i\ne \emptyset$.
						{From (\ref{equaltozero}), the existence of a $X[{\mathcal N}_{?i},\{i\}]\in {\mathbb C}^{|{\mathcal N}_{?i}|}$ making $\bar \Gamma X[:,\{i\}]=0$ for almost all $X[{\mathcal N}_{*i},\{i\}]\in {\mathbb C}^{|{\mathcal N}_{*i}|}$ implies
${\rm rank}(\bar \Gamma[:,{\mathcal N}_{?i}])={\rm rank}([\bar \Gamma[:,{\mathcal N}_{?i}],\bar \Gamma[:,{\mathcal N}_{*i}]\alpha])$ holds for almost all $\alpha\in {\mathbb C}^{|{\mathcal N}_{*i}|}$. By \cite[p.38]{Murota_Book}, for almost all $\alpha\in {\mathbb C}^{|{\mathcal N}_{*i}|}$, ${\rm rank}([\bar \Gamma[:,{\mathcal N}_{?i}],\bar \Gamma[:,{\mathcal N}_{*i}]\alpha])=\max_{\beta\in {\mathbb C}^{|{\mathcal N}_{*i}|}}{\rm rank}([\bar \Gamma[:,{\mathcal N}_{?i}],\bar \Gamma[:,{\mathcal N}_{*i}]\beta])$. Hence, it must hold that ${\rm Im}(\bar \Gamma[:,{\mathcal N}_{*i}])\subseteq {\rm Im}(\bar \Gamma[:, {\mathcal N}_{?i}])$.} However, this condition cannot be satisfied due to the existence of $x$ satisfying ${\rm Im} (x)\notin  {\rm Im} (\bar \Gamma[:,{\mathcal N}_{?i}])$, thus proving the necessity.
				\end{proof}}
				
				{\bf Proof of Theorem \ref{poly_verify}}:    Let $\bar \Gamma$ be defined as in the proof of Lemma \ref{theorem_k_less}. From the definition of ${\mathcal S}_i$, it is not difficult to see that the condition in Lemma \ref{theorem_k_less} holds if and only if ${\rm rank} (\bar \Gamma[:, {\mathcal N}_{?i}])={\rm rank} (\bar \Gamma[:,{\mathcal N}_i])(=\rho_i)$. Observe that ${\mathcal B}^*$ is the basis family of the dual of the matroid formed by the linear independence structure of rows of ${\mathcal M}[:,{\mathcal I}]$ (see \cite[Sec 2.3.2]{Murota_Book} for more details about {\emph{matroid and its dual}}). {Then, by the formula (2.64) in \cite[p.75]{Murota_Book}, for any ${\mathcal K}\subseteq {\mathcal J}_n$, $\max_{w\in {\mathcal B}^*}|w\cup {\mathcal K}|=|{\mathcal K}|+ {\rm grank} ({\mathcal M}[{\mathcal J}_n\backslash {\mathcal K},{\mathcal I}])-{\rm grank} ({\mathcal M}[{\mathcal J}_n,{\mathcal I}])$. Therefore, it generically  holds that
				${\rm rank} (\bar \Gamma[:,{\mathcal N}_i])= \max_{w\in {\mathcal B}^*}|w\cup {\mathcal {\mathcal N}_i}|=|{\mathcal N}_i|+ {\rm grank} ({\mathcal M}[{\mathcal J}_n\backslash {\mathcal N}_i,{\mathcal I}])-{\rm grank} ({\mathcal M}[:,{\mathcal I}])$, and similarly, ${\rm rank} (\bar \Gamma[:,{\mathcal N}_{?i}])=|{\mathcal N}_{?i}|+ {\rm grank} ({\mathcal M}[{\mathcal J}_n\backslash {\mathcal N}_{?i},{\mathcal I}])-{\rm grank}( {\mathcal M}[:,{\mathcal I}])$.}
				Immediately, (\ref{formula-nonempty}) follows.  $\hfill\square$
				

				\section{Conclusions} \label{sec-conclusion} 
				In this paper, we provide combinatorial insight into the rank resilience of pattern matrices from a generic viewpoint, in which instead of completely removing a set of nonzero entries, their values can be altered by structured perturbations.  Drawing upon tools from algebraic geometry, we establish the underlying generic properties in the addressed rank resilience problem, which can be extended to show the genericity of low-rank matrix pencil completion. Based on a basis preservation principle, we provide combinatorial conditions that are necessary and sufficient for the completability of GLRM$({\mathcal M},1)$. For the general GLRM$({\mathcal M},k)$ where $k>1$, we derive a necessary condition and a sufficient condition. As applications, our methods can justify and characterize the generic properties involving controllability and ISO of structured descriptor systems subject to structured perturbations, as well as SFM. Our results also contribute to LRMC by considering that the completions exhibit a prior zero-nonzero pattern and addressing it from a generic standpoint.

				Some interesting open problems could be future research:
				\begin{itemize}
					\item How to generalize the $k$-GRR framework to matrices with special structures, such as linear dependence, symmetry, positive semi-definiteness, Euclidean-distance matrix, and graph Laplacian matrix?
					\item Find the necessary and sufficient deterministic conditions for the completability of GLRM$({\mathcal M},k)$ ($k>1$).
					\item Extend the $k$-GRR framework to the generic low-rank {\emph{matrix pencil}} completion problem arising in systems and control (see Section \ref{application}).
				\end{itemize}

				\section*{Acknowledgements}
				The authors thank Prof. Dong Shijie for discussions on algebraic geometry theory {and Mr. Hongwei Zhang and Prof. Li Dai for fruitful discussions leading to an early draft of this work (version 1 of \cite{arxiv-version}) from the LRMC perspective.} 
}}

				{\bibliographystyle{ieeetr}
					{\tiny
						\bibliography{yuanz3}
				}}

				\end{document}